\documentclass[useAMS,usenatbib]{mn2e}
\usepackage{amsmath}
\usepackage{amsfonts}
\usepackage{array}
\usepackage{caption}
\usepackage{epsfig}
\usepackage{float}
\usepackage{gensymb}
\usepackage{graphicx}
\usepackage{longtable}
\usepackage{marginnote}
\usepackage{mathptmx}
\usepackage{multirow}
\usepackage{pdflscape}
\usepackage{placeins}
\usepackage{rotating}
\usepackage{subcaption}
\usepackage[table]{xcolor}
\usepackage{tabularx}
\usepackage{threeparttable}
\usepackage{wrapfig}
\usepackage{xfrac}
\usepackage{xfrac}

\def\reff@jnl#1{{\rm#1\/}}

\def\aj{\reff@jnl{AJ}}                  
\def\araa{\reff@jnl{ARA\&A}}            
\def\apj{\reff@jnl{ApJ}}                        
\def\apjl{\reff@jnl{ApJ}}               
\def\apjs{\reff@jnl{ApJS}}              
\def\ao{\reff@jnl{Appl.Optics}}         
\def\apss{\reff@jnl{Ap\&SS}}            
\def\aap{\reff@jnl{A\&A}}                       
\def\apjl{\reff@jnl{ApJ}}               
\def\aapr{\reff@jnl{A\&A~Rev.}}         
\def\aaps{\reff@jnl{A\&AS}}             
\def\azh{\reff@jnl{AZh}}                        
\def\baas{\reff@jnl{BAAS}}              
\def\jrasc{\reff@jnl{JRASC}}            
\def\memras{\reff@jnl{MmRAS}}           
\def\mnras{\reff@jnl{MNRAS}}            
\def\pra{\reff@jnl{Phys. Rev. A}}         
\def\prb{\reff@jnl{Phys. Rev. B}}         
\def\prc{\reff@jnl{Phys. Rev. C}}         
\def\prd{\reff@jnl{Phys. Rev. D}}         
\def\prl{\reff@jnl{Phys. Rev. Lett}}      
\def\pasp{\reff@jnl{PASP}}              
\def\pasj{\reff@jnl{PASJ}}              
\def\qjras{\reff@jnl{QJRAS}}            
\def\skytel{\reff@jnl{S\&T}}            
\def\solphys{\reff@jnl{Solar~Phys.}}    
\def\sovast{\reff@jnl{Soviet~Ast.}}     
\def\ssr{\reff@jnl{Space~Sci.Rev.}}     
\def\zap{\reff@jnl{ZAp}}                        
\def\nat{\reff@jnl{Nature}}             

\def\p#1by#2{{\partial{#1} \over \partial{#2}}}
\def\pp#1by#2#3{{\partial^2{#1} \over \partial{#2}\partial{#3}}}
\def\d#1by#2{{{\rm d}{#1} \over {\rm d}{#2}}}
\def\dd#1by#2#3{{{\rm d}^2{#1} \over {\rm d}{#2}{\rm d}{#3}}}

\title[A Study of Abell 3667 with KAT-7]{Early Science with the Karoo Array Telescope: a Mini-Halo Candidate in Galaxy Cluster Abell 3667}

\label{firstpage}

\author[C.~J.~Riseley et al.]{
 C.~J. ~Riseley$^{1}$\thanks{Corresponding author email: C.J.Riseley@soton.ac.uk}, 
 A.~M.~M.~ Scaife$^{1}$,
 N.~ Oozeer$^{2,3,4}$, L.~ Magnus$^2$ \& M.~W.~Wise$^{5,6}$
 \vspace{0.03in}\\
$^1$ School of Physics \& Astronomy, University of Southampton, Highfield, Southampton, SO17 1BJ, UK.\\
$^2$SKA South Africa, The Park, Park Road, Pinelands, Cape Town 7405, South Africa. \\
$^3$African Institute for Mathematical Sciences, 6-8 Melrose Road, Muizenberg 7945, South Africa.\\
$^4$Centre for Space Research, North-West University, Potchefstroom 2520, South Africa.\\
$^5$Netherlands Institute for Radio Astronomy (ASTRON), Postbus 2, 7990 AA Dwingeloo, The Netherlands. \\
$^6$Astronomical Institute ÒAnton PannekoekÓ, University of Amsterdam, Postbus 94249, 1090 GE Amsterdam, The Netherlands.
}

\date{Accepted ---; received ---; in original form \today}

\pagerange{\pageref{firstpage}--\pageref{lastpage}}

\pubyear{2014}

\begin{document}
\maketitle

\begin{abstract}
Abell 3667 is among the most well-studied galaxy clusters in the Southern Hemisphere. It is known to host two giant radio relics and a head-tail radio galaxy as the brightest cluster galaxy. Recent work has suggested the additional presence of a bridge of diffuse synchrotron emission connecting the North-Western radio relic with the cluster centre. In this work, we present full-polarization observations of Abell 3667 conducted with the Karoo Array Telescope at 1.33 and 1.82 GHz. Our results show both radio relics as well as the brightest cluster galaxy. We use ancillary higher-resolution data to subtract the emission from this galaxy, revealing a localised excess, which we tentatively identify as a radio mini-halo. This mini-halo candidate has an integrated flux density of $67.2\pm4.9$ mJy beam$^{-1}$ at 1.37 GHz, corresponding to a radio power of P$_{\rm{1.4\,\,\,GHz}}=4.28\pm0.31\times10^{23}$ W Hz$^{-1}$, consistent with established trends in mini-halo power scaling.
\end{abstract}


\section{Introduction}
Galaxy clusters are among the largest gravitationally-bound structures in the Universe. They form following the density profile of the dark matter distribution, from the merger of smaller galaxy groups into larger structures. As these structures grow, they undergo many dynamical interactions which make them excellent physical laboratories. Clusters grow through both accretion of gas from their environment as well as merger events with other galaxy groups and clusters. Major mergers are dramatic, releasing tremendous amounts of energy ($\sim10^{64}$ ergs; \citealt{2008SSRv..134...93F}) into the intra-cluster medium (ICM). One tell-tale indicator of clusters that have undergone recent merger events is the presence of large-scale, diffuse radio emission, examples of which are radio haloes and radio relics. Haloes are usually located in the cluster centre, whereas relics are typically observed toward the periphery at megaparsec distances from the cluster centre. Approximately 42 haloes and 36 relics\footnote{Counting only relics classified as `elongated' per the nomenclature of \protect{\cite{2012A&ARv..20...54F}}; these are analogous to the `radio gischt' of \protect{\cite{2004rcfg.proc..335K}}. We exclude `roundish' relics as these are more akin to the radio `phoenix' classification of source.} are currently known to be hosted by galaxy clusters (\citealt{2012A&ARv..20...54F}).

Common characteristics of radio relics include low surface-brightness (typically $\sim$1 $\mu$Jy arcsec$^{-2}$ at 1.4 GHz) and a steep spectral profile of the form $S(\nu) \propto \nu^{\alpha}$, where $\alpha$ is the spectral index taking typical values of $\alpha < -1$. The emission from such structures is synchrotron, indicative the presence of large-scale magnetic fields and relativistic electrons in the intracluster medium (ICM, see \citealt{2005AdSpR..36..729F}; \citealt{2008SSRv..134...93F}). Observations have shown that relics are highly-polarized \citep{2012A&ARv..20...54F} making them good targets for studying the physics of the magnetised ICM, as the magnetic field strength can be reconstructed by, for example, rotation measure (RM) synthesis (\citealt{2005A&A...441.1217B}) or equipartition arguments (\citealt{2005AN....326..414B}).

Observational data on relics is broadly consistent with an interpretation of the diffusive shock acceleration (DSA) model of electron acceleration (\citealt{2012A&ARv..20...54F}, and references therein). DSA models predict that the magnetic field within radio relics will be aligned with the shock front, and observations concur with this prediction (e.g. \citealt{2010Sci...330..347V}; \citealt{2012A&A...546A.124V}). Shocks have been confirmed using X-ray data (through detected surface brightness discontinuity and/or temperature jumps) in only a handful of clusters to-date, including for example Abell 754 \citep{2011ApJ...728...82M}, Abell 3376 \citep{2012PASJ...64...67A}, and Abell 3667 (e.g. \citealt{2010ApJ...715.1143F}; \citealt{2012PASJ...64...49A}). Typically observed shock Mach numbers lie in the range $1-3$; from simulations we expect a Mach number approximately in the range $2-3$ (\citealt{2012A&ARv..20...54F}, and references therein).

Relics are also known to be hosted by some cool-core clusters which suggests both major and minor/off-axis merger events can be energetic enough to generate relics. Although they exist on vastly different scales, examples of this are the small-scale (a few $\times100$kpc in extent) relic in Abell 85 (\citealt{2001AJ....122.1172S}, \citealt{2014AJ....148...23S}) and the giant ($\sim2$ Mpc size) radio relic in the merging cluster Abell 115 (\citealt{2001A&A...369..441G}, \citealt{2005ApJ...619..161G}). Additionally, X-ray observations of Abell 115 suggest that while the (off-axis) merger event has significantly disturbed the morphology of the two subclusters, both possess intact cool cores \citep{2005ApJ...619..161G}.

Additionally, some cool-core clusters are known to host smaller-scale diffuse radio emission known as mini-haloes. Radio mini-haloes (MH) are diffuse radio sources on a smaller scale than the giant radio haloes (GRH) and relics seen in dynamically-active clusters (typical MH scale size $\leq500$kpc; GRH scale size $\sim1$Mpc). MH are typically co-located with the brightest cluster galaxy (BCG) and although few are known (15 confirmed; \citealt{2014ApJ...781....9G}) a general scaling trend is observed between BCG radio power and MH radio power \citep{2014ApJ...781....9G}. MH typically exhibit spectra that are slightly steeper than those of relics or giant haloes ($\alpha_{\rm{MH}} < -1.2$). Two main models exist for the mechanism responsible for MH electron acceleration: magnetohydrodynamic (MHD) turbulence within the cool core and secondary electron (or `hadronic') models.

Two principal sources of MHD turbulence have been suggested: (1) turbulence generated by a cooling flow (for example \citealt{2002A&A...386..456G}) and (2) turbulence induced by gas sloshing within the cool core \citep{2011MmSAI..82..632Z}. Evidence for each source of turbulence is provided by, for example, (1) the correlation between cooling flow power and MH power \citep{2004A&A...417....1G} and (2) the detection of spiral-shaped cold fronts in observations (for example \citealt{2009ApJ...704.1349O}) and simulations \citep{2013ApJ...762...78Z}. The observed trends in MH/BCG co-location and radio power suggest a potential source of the electrons responsible for the MH radio emission (\citealt{2008A&A...486L..31C}; \citealt{2014ApJ...781....9G}).

The hadronic group of models predicts that the population of relativistic electrons responsible for MH arises from interactions between cosmic ray protons (CRp) and protons of the ICM \citep{1980ApJ...239L..93D}. This group of models can reproduce some properties of MH, but purely-hadronic models generally fail to replicate a number of observed characteristics, including the radial steepening of the spectral index and the detected high-frequency cutoff in some MH spectra \citep{2014ApJ...781....9G}.

\subsection{Abell 3667}
Located at a redshift of $z=0.0553$ \citep{2009ApJ...693..901O} Abell 3667 (hereinafter A3667) is a massive double-cluster of galaxies with $\textrm{M}_{\textrm{gas}}>10^{15}\,\,\,\textrm{M}_{\odot}$ \citep{1992MNRAS.259..233S}. It is believed to have undergone a recent major merger \citep{2010ApJ...715.1143F} and is known to host two large-scale, diffuse radio relics (for example \citealt{1997MNRAS.290..577R}). The brightest cluster galaxy (BCG) is head-tail radio galaxy MRC B2007-569 (per the nomenclature of \citealt{1982MNRAS.198..259G}).

Optical observations have yielded evidence of substructure in A3667: the cluster is believed to consist of two main sub-clusters of 242 (164) cluster members in the primary (secondary) sub-cluster and one tertiary sub-group with 27 members (Owers et al. 2009). The primary sub-cluster density distribution lies coincident with the peak of X-ray emission; the secondary sub-cluster is to the North-West of the cluster centre and the tertiary is to the South-East. Optical observations also suggest that the merger event is occurring close to the plane of the sky (\citealt{2008A&A...479....1J}, Owers et al. 2009).

A3667 hosts two giant radio relics (\citealt{1997MNRAS.290..577R}; \citealt{2003PhDT.........3J}) North-West (NW) and South-East (SE) of the cluster centre. Observations show the NW relic has an integrated flux density in the range 2--4 Jy at 1.4 GHz (\citealt{1997MNRAS.290..577R}: 2.4$\pm$0.2 Jy; \citealt{2003PhDT.........3J}: 3.7$\pm$0.3 Jy) and a spectral index of approximately $-1.1$ between 1.4 and 2.4 GHz \citep{1997MNRAS.290..577R}. The SE relic has an integrated flux density of 0.30$\pm$0.02 \citep{2003PhDT.........3J}.

There is tentative evidence of a radio halo in A3667, with an integrated flux density of 33$\pm6$ mJy at 1.4 GHz (ATCA: \citealt{2003PhDT.........3J}) and 44$\pm$6 mJy at 3.3 GHz (Parkes: \citealt{2013MNRAS.430.1414C}). However, due to the difference between angular scales recovered by the two instruments, the recovered flux density at 1.4 GHz should be considered a lower limit only. In addition, recent single-dish work has suggested the presence of an unpolarized radio `bridge' of synchrotron emission (from electrons re-accelerated by turbulence in the post-shock region) connecting the NW radio relic with the cluster centre, with surface brightness $0.21-0.25 \, \, \mu\rm{Jy arcsec}^{-2}$ at 2.3 GHz \citep{2013MNRAS.430.1414C}.

From observations at radio wavelengths, the magnetic field strength in the NW relic has been estimated. RM analysis of two point sources background to the NW relic yields estimates of $3.0-5.1 \, \, \mu$G \citep{2004rcfg.proc...51J}; for comparison, equipartition arguments yield values of $1.5-2.5 \mu$G \citep{2004rcfg.proc...51J}. The magnetic field strength in A3667 has also been estimated from X-ray observations: lower limits of $1.6 \, \, \mu\rm{G}$ \citep{2009PASJ...61..339N} and $3.0 \, \,  \mu$G \citep{2010ApJ...715.1143F} in the NW relic; central magnetic field strength $7.0-16.0 \mu$G \citep{2001ApJ...549L..47V}. These estimates of the magnetic field strength are summarised in Table \ref{bfields}. 

Jumps in X-ray hardness and surface brightness are detected at the outer edge of the NW relic, consistent with predictions based on the shock (re-)acceleration model, with a Mach number of $M = 2.09\pm0.47$ and shock speed $v_{\rm{shock}} = 1210\pm220$ \citep{2010ApJ...715.1143F}. A front of cold gas is detected coincident with the BCG \citep{1999ApJ...521..526M}. \cite{2010A&A...513A..37H} apply 16 cool-core diagnostics to the Chandra HIghest X-ray FLUx Galaxy Cluster Sample (HIFLUGCS; \citealt{2002ApJ...567..716R}) to identify the best parameters for characterising cool-core clusters; following which A3667 is classified as a weak cool-core cluster.

Using the Murchison Widefield Array (MWA; \citealt{2013PASA...30....7T}) at 149 MHz, \cite{2014MNRAS.445..330H} recover an integrated flux density of $28.1\pm1.7$ ($2.4\pm0.1$) Jy for the NW (SE) relics, and an average spectral index of $\alpha=-0.9\pm0.1$ for both. From the spectral index of the NW relic, the data imply a shock Mach number of 2.4$\pm0.1$; within the range expected from DSA and consistent with the Mach number derived from X-ray observations by \cite{2010ApJ...715.1143F}. The NW relic exhibits a spectral index gradient that steepens toward the cluster centre, indicative of an ageing electron population; the radial profile is consistent with that of an expanding shock travelling out from the cluster centre, as has been observed in a number of other relics (for example the relic in the 'Sausage cluster'; \citealt{2010Sci...330..347V}). Additionally, Hindson et al. recover flux from both the radio bridge region and radio halo at lower significance than expected from previous work; the interpretation is that previous detections of a bridge and halo in A3667 are the result of poorly-sampled Galactic emission that contaminates the region.

\begin{table}
\begin{center}
\caption{Summary of previous magnetic field strength derivations for A3667. \label{bfields}}
\begin{tabular}{llrcc}
\hline\hline
 Paper & Estimation method & Region & B\hspace{0.3cm} \\
 & & & ($\, \, \mu\rm{G}$)\hspace{0.1cm} \\
\hline
Vikhlinin et. al. (2001) & X-Ray (Chandra) & Centre & 7.0-16.0\\
\cite{2004rcfg.proc...51J} & RM estimates & NW relic & 3.0-5.1 \\
\cite{2004rcfg.proc...51J} & Equipartition & NW relic & 1.5-2.5 \\
\cite{2009PASJ...61..339N} & X-Ray (Suzaku) & NW relic & $>1.6$\\
\cite{2010ApJ...715.1143F} & X-Ray (XMM) & NW relic & $>3.0$ \\ 
\hline\hline
\end{tabular}
\end{center}
\end{table}

In this work, A3667 was observed with the Karoo Array Telescope, a precursor to MeerKAT and the Square Kilometre Array, to investigate the nature of the diffuse radio emission from the relics and `bridge' of the cluster. This paper is structured as follows: we discuss the Karoo Array Telescope, the observations, and the reduction process in \S\ref{sec:KAT}. Our results are presented in \S\ref{sec:RES} and in \S\ref{sec:AN}, while the implications are discussed in \S\ref{sec:DISC}. We draw our conclusions in \S\ref{sec:CONC}. Throughout this work, we assume a concordance cosmology of H$_{0}=73$ km s$^{-1}$ Mpc$^{-1}$, $\Omega_{\rm{m}} = 0.27$, $\Omega_{\rm{vac}} = 0.73$. All errors are quoted to 1-$\sigma$. We adopt the spectral index convention that $S \propto \nu^{\alpha}$ and we take the redshift of A3667 to be 0.0553 (Owers et al. 2009). At this redshift, an angular distance of 1$^{\prime\prime}$ corresponds to 1.031 kpc \citep{2006PASP..118.1711W}.

\section{The Karoo Array Telescope} \label{sec:KAT}
The Karoo Array Telescope (KAT-7) is a synthesis telescope consisting of seven 12-m diameter dishes with baselines in the range $26 - 185$\,m, located in the Republic of South Africa. It has dual linear-feed receivers capable of observing in the range 1.20--1.95\,GHz, with an IF bandwidth of 400\,MHz. However, due to analog filters in the IF and baseband system, only the central 256\,MHz of the IF bandwidth is useable.

This work was conducted as part of KAT-7 commissioning; as such central observing frequencies were selected to be 128\,MHz from the bottom and top end of the observable bandwidth, yielding useable slices that cover the extremes of the KAT-7 frequency range. Our reference frequencies for this work are therefore 1382 and 1826 MHz. With a longest baseline of 819 (1124)\,$\lambda$ at 1328 (1822)\,MHz, KAT-7 has modest angular resolution of $\theta\simeq 4\,(3)$\,arcmin, beneficial to recovering diffuse low surface-brightness emission.

KAT-7 is capable of observing using a number of correlator modes (see http://public.ska.ac.za/kat-7 for further details). In this work, wideband mode was used, with the IF divided into 1024 channels; after removal of the bandpass edges, the central 601 channels were selected for a useable bandwidth of 235\,MHz. Table \ref{sensitivity} provides a summary of telescope characteristics. Although primarily a testbed for new materials and engineering for the upcoming MeerKAT and Square Kilometre Array (SKA) projects, early science is underway with KAT-7 (see for example \citealt{2013MNRAS.433.1951A}).

\begin{table}
\caption{Technical Details of KAT-7 \label{sensitivity}}
\begin{threeparttable}
\begin{tabular}{lrr}
\hline\hline
Number of Antennas & \multicolumn{2}{c}{7} \\
Dish Diameter (m) & \multicolumn{2}{c}{12.0} \\
\emph{uv}-min (m) & \multicolumn{2}{c}{25.6} \\
\emph{uv}-max (m) & \multicolumn{2}{c}{185.0} \\
System Temperature (K) & \multicolumn{2}{c}{30} \\
IF Bandwidth (MHz) & \multicolumn{2}{c}{400.0}\\
Useable Bandwidth $\Delta\rm{\nu}$ (MHz) & \multicolumn{2}{c}{256.0} \\
\emph{Effective Bandwidth in this Work} (MHz) & 157.0\tnote{\dag} & 235.0 \\
\emph{Effective Number of Channels in this Work} & 401\tnote{\dag} & 601 \\
Polarization & \multicolumn{2}{c}{XX,YY,XY,YX} \\
\hline
Reference Frequency $\rm{\nu_{ref}}$ (MHz) & 1328 & 1822 \\
\emph{Effective Reference Frequency} $\rm{\nu_{eff}}$ (MHz) & 1372\tnote{\dag} & 1826 \\
Resolution $\rm{\theta}$ (arcmin) & 4.19 & 3.06 \\
Largest detectable size $\rm{\theta}_{\rm{max}}$ (arcmin) & 30.31 & 22.09 \\

\hline\hline
\end{tabular}

\begin{tablenotes}
	\item[\dag] Reduced bandwidth due to RFI environment -- see \S\ref{sec:rfi} for further details.
\end{tablenotes}
\end{threeparttable}
\end{table}

\subsection{Observation Details}
\subsubsection{KAT-7}
\begin{figure}
	\begin{subfigure}[t]{0.45\textwidth}
		\includegraphics[width=\textwidth]{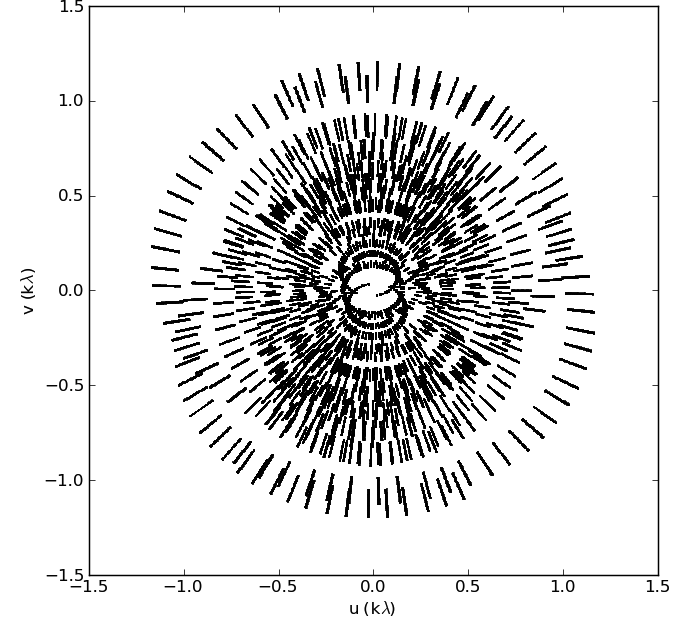}
		\label{fig:uvcovg_1826}
	\end{subfigure}
	\begin{subfigure}[b]{0.45\textwidth}
		{\includegraphics[width=\textwidth]{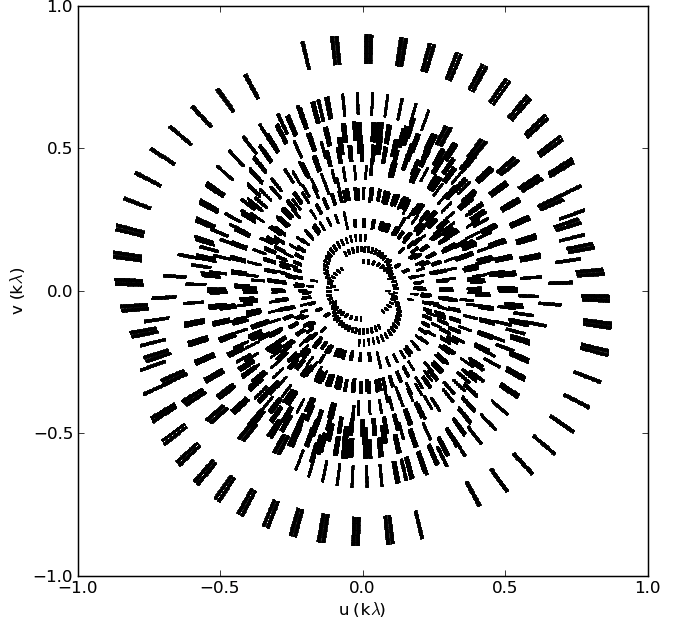}}
		\label{fig:uvcovg_1328}
	\end{subfigure}
	\caption{\emph{uv}-coverage of A3667. \textsc{top:} 1822 MHz, \textsc{bottom:} 1328 MHz. Both figures plot the \emph{uv}-coverage of the central pointing in the mosaic.}
	
\label{fig:a3667_uvcovg}
\end{figure}

A3667 was observed in full-polarization in both upper and lower bands. Observations at 1822 MHz were conducted in 2013 March, and 1328 MHz data was taken during 2013 May. Between these observing periods, KAT-7 was upgraded with new receivers to reduce intrinsic radio frequency interference (RFI). All results in this paper were produced with a single observing run at 1822 MHz, and two runs at 1328 MHz. Table \ref{pointingsummary} summarises the details of observations. At 1822 MHz, 9 close-packed pointings were used to cover the cluster region, while 7 pointings were used at 1328 MHz. The KAT-7 primary beam has full-width at half maximum (FWHM) of approximately $1\degree$ at 1822 MHz. The \emph{uv}-coverage of the central pointing on A3667 at each frequency is shown in Figure \ref{fig:a3667_uvcovg}.

During each observing run, three sources were used as calibrators: PKS B1934-638, 3C138/3C286 and PMN J2040-5735. PKS B1934-638 (hereafter 1934-638) was used as both a bandpass calibrator and flux calibrator, 3C138 (3C286) was the polarization calibrator at 1822 (1328) MHz, and PMN J2040-5735 was to be phase calibrator. However, the local RFI environment of PMN J2040-5735 proved too strong to derive good calibration solutions, so 1934-638 was substituted as phase calibrator, as it was interleaved sufficiently often to be suitable. In the lower band, 1934-638 was interleaved every 30 minutes. In the upper band, it was interleaved every 60 minutes. Phase solutions in both bands were smoothly-varying, with less than $10\degree$ variation over the entire observation, and variation of less than 3$\degree$ between scans.

\begin{table*}
\begin{center}
\caption{KAT-7 pointings on A3667 at 1822 MHz (ACO3667-H*) and 1328 MHz (ACO3667-L*), with their individual integration times and both expected thermal noise component and measured rms based on individual images made for diagnostic purposes. \label{pointingsummary}}
\begin{tabular}{cccccc}
\hline\hline
                     &                &                &                               & Thermal                             & Measured   \\ 
Pointing ID & RA         & Dec        & $\tau_{\rm{int}}$ & $S_{\rm{rms}}$                 & $S_{\rm{rms}}$\\
                     & (J2000) & (J2000) & (s)                       & ($\mu$Jy beam$^{-1}$)    & (mJy beam$^{-1}$) \\
\hline
ACO3667-H01 & 20$\textsuperscript{h}$11$\textsuperscript{m}$16.$\textsuperscript{s}$60 & -56\textdegree59$^{\prime}$57$^{\prime\prime}$.00 & 4110 & 61.18 & 1.38 \\ 
ACO3667-H02 & 20$\textsuperscript{h}$11$\textsuperscript{m}$16.$\textsuperscript{s}$90 & -56\textdegree49$^{\prime}$57$^{\prime\prime}$.00 & 4280 & 59.96 & 1.35 \\ 
ACO3667-H03 & 20$\textsuperscript{h}$11$\textsuperscript{m}$17.$\textsuperscript{s}$19 & -56\textdegree39$^{\prime}$57$^{\prime\prime}$.00 & 4230 & 60.31 & 1.43 \\ 
ACO3667-H04 & 20$\textsuperscript{h}$12$\textsuperscript{m}$29.$\textsuperscript{s}$99 & -56\textdegree59$^{\prime}$58$^{\prime\prime}$.99 & 4210 & 60.45 & 0.94 \\ 
ACO3667-H05 & 20$\textsuperscript{h}$12$\textsuperscript{m}$29.$\textsuperscript{s}$99 & -56\textdegree49$^{\prime}$58$^{\prime\prime}$.99 & 4250 & 60.17 & 1.14 \\
ACO3667-H06 & 20$\textsuperscript{h}$12$\textsuperscript{m}$29.$\textsuperscript{s}$99 & -56\textdegree39$^{\prime}$58$^{\prime\prime}$.99 & 4310 & 59.75 & 1.60 \\ 
ACO3667-H07 & 20$\textsuperscript{h}$13$\textsuperscript{m}$43.$\textsuperscript{s}$39 & -56\textdegree59$^{\prime}$57$^{\prime\prime}$.00 & 4240 & 60.24 & 1.52 \\
ACO3667-H08 & 20$\textsuperscript{h}$13$\textsuperscript{m}$43.$\textsuperscript{s}$10 & -56\textdegree49$^{\prime}$57$^{\prime\prime}$.00 & 4130 & 61.04 & 1.18 \\
ACO3667-H09 & 20$\textsuperscript{h}$13$\textsuperscript{m}$42.$\textsuperscript{s}$80 & -56\textdegree39$^{\prime}$57$^{\prime\prime}$.00 & 4190 & 60.60 & 1.60 \\
\hline
ACO3667-L01 & 20$\textsuperscript{h}$12$\textsuperscript{m}$06.$\textsuperscript{s}$13 & -56\textdegree49$^{\prime}$12.$^{\prime\prime}$30 & 5380 & 67.23 & 2.53 \\
ACO3667-L02 & 20$\textsuperscript{h}$11$\textsuperscript{m}$02.$\textsuperscript{s}$14 & -56\textdegree49$^{\prime}$12.$^{\prime\prime}$30 & 5550 & 66.19 & 1.99 \\
ACO3667-L03 & 20$\textsuperscript{h}$11$\textsuperscript{m}$34.$\textsuperscript{s}$14 & -56\textdegree35$^{\prime}$20.$^{\prime\prime}$90 & 5480 & 66.61 & 1.85 \\
ACO3667-L04 & 20$\textsuperscript{h}$12$\textsuperscript{m}$29.$\textsuperscript{s}$56 & -56\textdegree35$^{\prime}$20.$^{\prime\prime}$90 & 5310 & 67.67 & 2.79 \\
ACO3667-L05 & 20$\textsuperscript{h}$13$\textsuperscript{m}$10.$\textsuperscript{s}$14 & -56\textdegree49$^{\prime}$12.$^{\prime\prime}$30 & 5360 & 67.35 & 2.38 \\
ACO3667-L06 & 20$\textsuperscript{h}$12$\textsuperscript{m}$29.$\textsuperscript{s}$56 & -57\textdegree03$^{\prime}$03.$^{\prime\prime}$70 & 5420 & 66.98 & 1.93 \\
ACO3667-L07 & 20$\textsuperscript{h}$11$\textsuperscript{m}$02.$\textsuperscript{s}$14 & -57\textdegree03$^{\prime}$03.$^{\prime\prime}$70 & 5220 & 68.25 & 2.54 \\
 \hline\hline
\end{tabular}
\end{center}
\end{table*}

It has been suggested that 3C138 exhibits variation in its polarized flux emission, resulting from a flare that began in 2003 \citep{2013ApJS..206...16P} so 3C286 was chosen as polarization calibrator at 1328 MHz. The polarization calibrator was observed for 4 scans during each observing run, for a total integration time of order 3000s.

\subsection{Reduction Process}

\subsubsection{RFI Excision}\label{sec:rfi}
Initial pre-processing was performed on the data with an in-house automatic RFI flagging pipeline (priv. comm., Mauch~T.). Subsequently, the data were visually inspected to remove strong sources of RFI, and then automatically flagged using \texttt{flagdata}\footnote{http://casaguides.nrao.edu/index.php?title=Flagdata} in \textsc{casa 4.1}\footnote{http://casa.nrao.edu}.

For the upper band ($\nu_{\rm{ref}}=1822$ MHz) observations, KAT-7 was equipped with its first-generation receivers; both low-level and sharp RFI peaks were present in the data. Additionally, the first-generation receivers experienced strong, broad-band internal RFI -- an issue which has since been corrected with the upgrade to the second-generation receivers. Following visual inspection approximately 50 per cent of the data was flagged out across the entire bandwidth, resulting in an effective reference frequency of 1826 MHz.

Prior to the lower band ($\nu_{\rm{ref}}=1328$ MHz) observations being conducted, KAT-7 was upgraded with improved second-generation receivers to eliminate instrumental RFI. In the lower band highly-significant RFI from a known source dominated the signal below approximately 1280\,MHz. Hence a cut was made at 1294 MHz, resulting in an effective bandwidth of 157 MHz and effective reference frequency (adopted henceforth) of 1372 MHz. Aside from this cut, approximately 16 per cent of the data was flagged across the bandwidth. These effective reference frequencies and effective bandwidths are also summarised in Table \ref{sensitivity}.

\subsubsection{Calibration}
The results presented in this work have been calibrated using \textsc{casa 4.1}. The flux scale was set using the primary calibrator (1934-638) on the Perley-Butler (2010) model which yields a Stokes I flux density of 14.795 Jy at 1450 MHz and 13.159 Jy at 1944 MHz. Parallel-hand delays, complex gains and bandpass solutions were derived for each antenna using 1934-638, following standard practice. For calibration, 1934-638 was assumed to have zero polarization; although 1934-638 is known to have no linear polarization, it does have low-level circularly-polarized emission ($\sim0.03$ per cent; \citealt{2000MNRAS.319..484R}). This assumption of zero polarization will lead to a slight offset in Stokes V, which should be corrected if analysis of that parameter is to be made. However, in this work we neglect this offset.

Cross-hand delays and phase solutions were determined using the polarization calibrator (3C286 at 1328 MHz; 3C138 at 1822 MHz) assuming a non-zero polarization model. Stokes parameters were derived directly from the measured complex visibilities and then compared with the source model from \cite{2013ApJS..206...16P} (see Table \ref{tab:polcalmodels}) in order to correct for the 180$\degree$ ambiguity present in the XY phase solution\footnote{We note that this ambiguity can be avoided by use of an injected XY phase calibration signal, such as that utilised by the ATCA instrument.}. With these additional calibration solutions, the complex gains were solved for the polarization calibrator using the full-Stokes source model from \cite{2013ApJS..206...16P} and used to calibrate the polarization leakage. Typical leakages were of the order of 3 per cent in the upper band and 3--6 per cent in the lower band.

From these data we recover a polarization fraction of $8.4\pm0.6$ per cent and electric vector position angle (EVPA) of $31.8\pm0.7\degree$ for 3C286; for 3C138 the polarization fraction is $8.1\pm0.6$ per cent and EVPA $-11.0\pm0.2\degree$, in reasonable agreement with those determined by \cite{2013ApJS..206...16P} for these sources. See Table \ref{tab:polcalmodels} for details.

\begin{table}
\begin{center}
\caption{Polarization properties of 3C286 and 3C138 from \protect\cite{2013ApJS..206...16P} and those derived from data taken with KAT-7 during these observations. \label{tab:polcalmodels}}
\begin{tabular}{llccc}
\hline\hline
Source & & Freq. & Fractional Pol. & EVPA $\chi$ \\
              & & (MHz) & ($\%$) & ($\degree$) \\
\hline
\multirow{4}{*}{3C286} & \multirow{4}{*}{Model} & 1050 & 8.6 & 33 \\
					& & 1450 & 9.5 & 33 \\
					& & 1640 & 9.9 & 33 \\
					& & 1950 & 10.1 & 33 \\\cline{2-5}

					& Recovered & 1372 & $8.4\pm0.6$ & $31.8\pm0.7$ \\ 
\hline
\multirow{4}{*}{3C138} & \multirow{4}{*}{Model} & 1050 & 5.6 & $-14$ \\
					& & 1450 & 7.5 & $-11$ \\
					& & 1640 & 8.4 & $-10$ \\
					& & 1950 & 9.0 & $-10$ \\\cline{2-5}

					& Recovered & 1826 & $8.2\pm0.6$ & $-11.0\pm0.2$ \\ 
\hline\hline
\end{tabular}
\end{center}
\end{table}

\subsubsection{Imaging}
Deconvolution was performed in all Stokes parameters using multi-scale \texttt{clean} in \textsc{casa}. The deconvolution process used natural weighting, with scales set to [0 (point sources), 1$\times$beam, 2$\times$beam] to improve sensitivity to diffuse emission. Maps were produced at reference frequencies of 1826 and 1372 MHz. The fields were mosaicked during deconvolution, cleaning initially down to a threshold of 1.5$\sigma_{\rm{c}}$ (where $\sigma_{\rm{c}}$ is the theoretical confusion limit -- see section \S\ref{sec:conf}) before inspection. Subsequently, further cleaning was performed down to a threshold of $\sigma_{\rm{c}}$. Finally, the images were corrected for the primary beam response. In both cases, the cluster was imaged out to 10 per cent of the primary beam response. See Figures \ref{fig:1826_StokesI}, \ref{fig:1372_StokesI} and \ref{fig:linpolmaps} for the maps.

\subsection{Confusion and Thermal Noise}\label{sec:conf}
At the angular resolution of KAT-7, we expect our results to be limited by confusion noise rather than thermal noise in Stokes I. KAT-7 has a system temperature of $\rm{T}_{\rm{sys}} = 30$ K and we expect thermal noise values of $\sigma_{\rm{rms}} = 20.13 \,\, (25.39) \,\, \rm{\mu}$Jy $\rm{beam}^{-1}$ at 1822 (1328) MHz for the full observation. Pointings at each frequency, integration times and both predicted thermal noise and measured rms per pointing are listed in Table \ref{pointingsummary}. The given rms values are measured from images of each pointing made individually for diagnostic purposes.

We derive the confusion limit using a 1.4 GHz differential source counts model which unifies contributions from flat-spectrum radio quasars, BL Lac objects and steep-spectrum objects \citep{2010A&ARv..18....1D}. Defining confusion as when a level of one source per 20 beam areas is reached, we interpolate this 1.4 GHz model to find the corresponding flux density cutoff for confusion
\begin{equation}\label{eq:conf_n}
N = \int_{S_{0}}^\infty \Omega \,\,\,\, n(S) \,\, dS 
\end{equation}
where $N$ is the number of sources per beam area, $S_{0}$ is the flux density cutoff, $\Omega$ is the beam area in steradians and $n(S)$ is the differential source counts model (in Jy$^{-1}$ sr$^{-1}$). We subsequently use $S_{0}$ to integrate the differential source counts model to derive the confusion limit
\begin{equation}\label{eq:conf_lim}
\sigma_{\rm{c}}^2 = \int_0^{S_{0}} \Omega \,\,\,\, n(S) \,\, S^2 \,\, dS 
\end{equation}
where $\sigma_{\rm{c}}$ is the confusion limit and $S$ is the source flux density. KAT-7 has beam size 3.05$^{\prime}$ (4.19$^{\prime}$) at 1822 (1328) MHz, therefore we derive values of $\sigma_{\rm{c}}$ = 1.28 (2.15) mJy beam$^{-1}$. These confusion limit values dominate over the predicted thermal noise. We also note that KAT-7 rapidly becomes confusion-limited, as the thermal noise drops below the confusion limit after $\tau_{\rm{int}} \sim 8$s at 1822 MHz and $\tau_{\rm{int}} \sim 4$s at 1328 MHz.

\subsection{Archival ATCA Data}
Archival Australia Telescope Compact Array (ATCA) data was used to provide a model of the head-tail radio galaxy MRC B2007-569 (\citealt{1982MNRAS.198..259G}; hereafter B2007-569) for subtraction purposes. B2007-569 was observed with the ATCA in 1.5D configuration, with baselines in the range 107\,m -- 4.4\,km, providing a nominal resolution of 10 arcsec and maximum detectable scale size of 7 arcmin at 1.4 GHz. Observations were taken on 2011 November 26 (project code CX224, P.I. Carretti E.) for a total integration time of 9\,h.

The data were calibrated with the Multichannel Image Reconstruction, Image Analysis and Display (\textsc{miriad}; \citealt{1995ASPC...77..433S}) package following standard techniques. The flux density scale was set using the Reynolds (1994) model for 1934-638, which yields flux densities consistent with those set by the \textsc{casa} task \texttt{setjy}. Therefore we can be confident that both the KAT-7 and ATCA observations will have consistent flux scaling. Following calibration, the full 2\,GHz Compact Array Broadband Backend (CABB; \citealt{1994ApJS...91..111W}) bandwidth was divided into subbands of 200\,MHz width. The calibrated data were imported into \textsc{casa} for further flagging and initial imaging using the task \texttt{clean}, followed by 5 rounds of phase-only self-calibration and imaging. Images were made in all subbands above 1180\,MHz to enable modelling of the spectral variation of B2007-569. The 1180\,MHz band was excluded as it was severely flagged (due to RFI excision and edge channel removal) resulting in a useable frequency coverage of 1288 -- 3124 MHz.

To obtain a clean-component model (CCM) for subtraction from our KAT-7 data, we performed multi-frequency synthesis using the \textsc{msmfs} algorithm \citep{2011A&A...532A..71R} as implemented in the \textsc{casa} task \texttt{clean}. This algorithm uses the large fractional bandwidth available to deconvolve the image using multiple spatial scales and Taylor coefficients (as a function of frequency) to describe the intensity distribution. After deconvolution, a spectral index map and its associated error map are produced. We performed \texttt{mfs}-\texttt{clean} with the reference frequency set to 1400\,MHz in order to provide a model for subtraction from the KAT-7 data most sensitive to large-scale structure. See \S\ref{sec:sourcesub} for further details on the subtraction, maps and analysis.

\section{Results}\label{sec:RES}
\begin{figure*}
\hspace{0.4cm}
\includegraphics[width=0.73\textwidth]{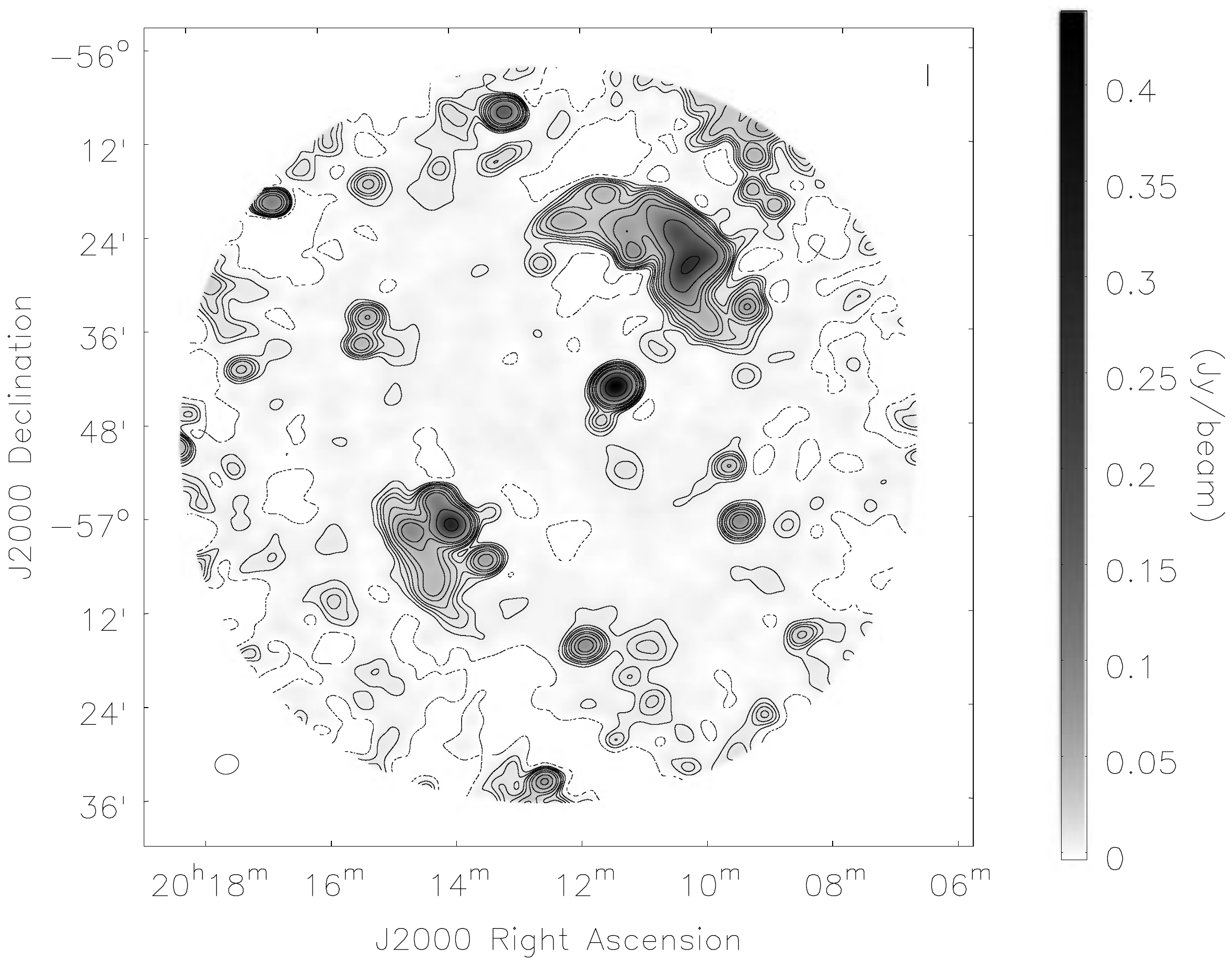}
\caption{Total intensity map of A3667 at 1826 MHz. Image rms is 1.30 mJy beam$^{-1}$. Contours mark [-1, 1, 2, 3, 4, 6, 8, 12, 16, 32, 64, 128] x $3\sigma_{\rm{rms}}$. Beam size (unfilled circle, bottom-left) is $184\times150$ arcsec.}
\label{fig:1826_StokesI}
\end{figure*}

\begin{figure*}
\centering
\includegraphics[width=0.73\textwidth]{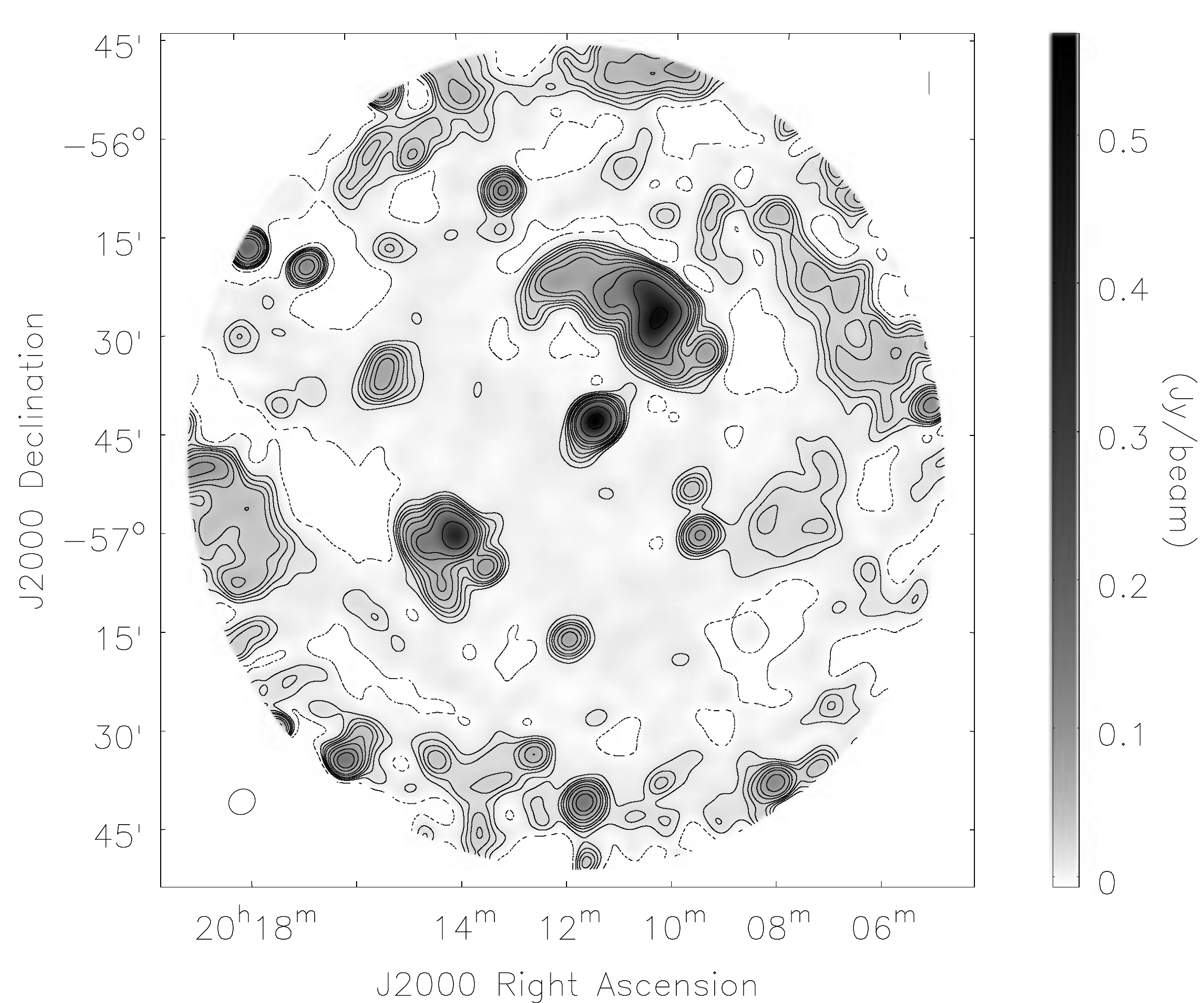}
\caption{Total intensity map of A3667 at 1372 MHz. Image rms is 2.18 mJy beam$^{-1}$. Contours mark [-1, 1, 2, 3, 4, 6, 8, 12, 16, 32, 64, 128] x $3\sigma_{\rm{rms}}$. Beam size (unfilled circle, bottom-left) is $250\times223$ arcsec.}
\label{fig:1372_StokesI}
\end{figure*}

We present total intensity (Stokes I) images at 1826 MHz (Figure \ref{fig:1826_StokesI}) and 1372 MHz (Figure \ref{fig:1372_StokesI}). Throughout all figures, total intensity contours mark [-1, 1, 2, 3, 4, 6, 8, 12, 16, 32, 64, 128] x $3\sigma_{\rm{rms}}$ unless otherwise stated. 

Figure \ref{fig:1826_StokesI} has image noise $\sigma_{\rm{rms}} = 1.30$ mJy beam$^{-1}$, a value comparable to the confusion limit at this frequency (see \S\ref{sec:conf}). Large-scale diffuse emission from both radio relics is clearly detected, as are many point sources in the field of view, and the head-tail radio galaxy B2007-569. With the resolution of KAT-7, we do not resolve the structure of B2007-569. The NW relic extends for an angular size of 33 arcmin, corresponding to a physical distance of 2.04 Mpc at this redshift; the SE radio relic covers an extent of 23 arcmin or 1.42 Mpc. Figure \ref{fig:1372_StokesI} has image noise level $\sigma_{\rm{rms}} =  2.18$ mJy beam$^{-1}$, comparable to the confusion limit derived in \S\ref{sec:conf}. With the marginally-lower resolution, the diffuse sources exhibit smoother morphologies than in Figure \ref{fig:1826_StokesI}.

\begin{figure*}
	\hspace{0.6cm}
	\begin{subfigure}[t]{0.745\textwidth}
		\includegraphics[width=\textwidth]{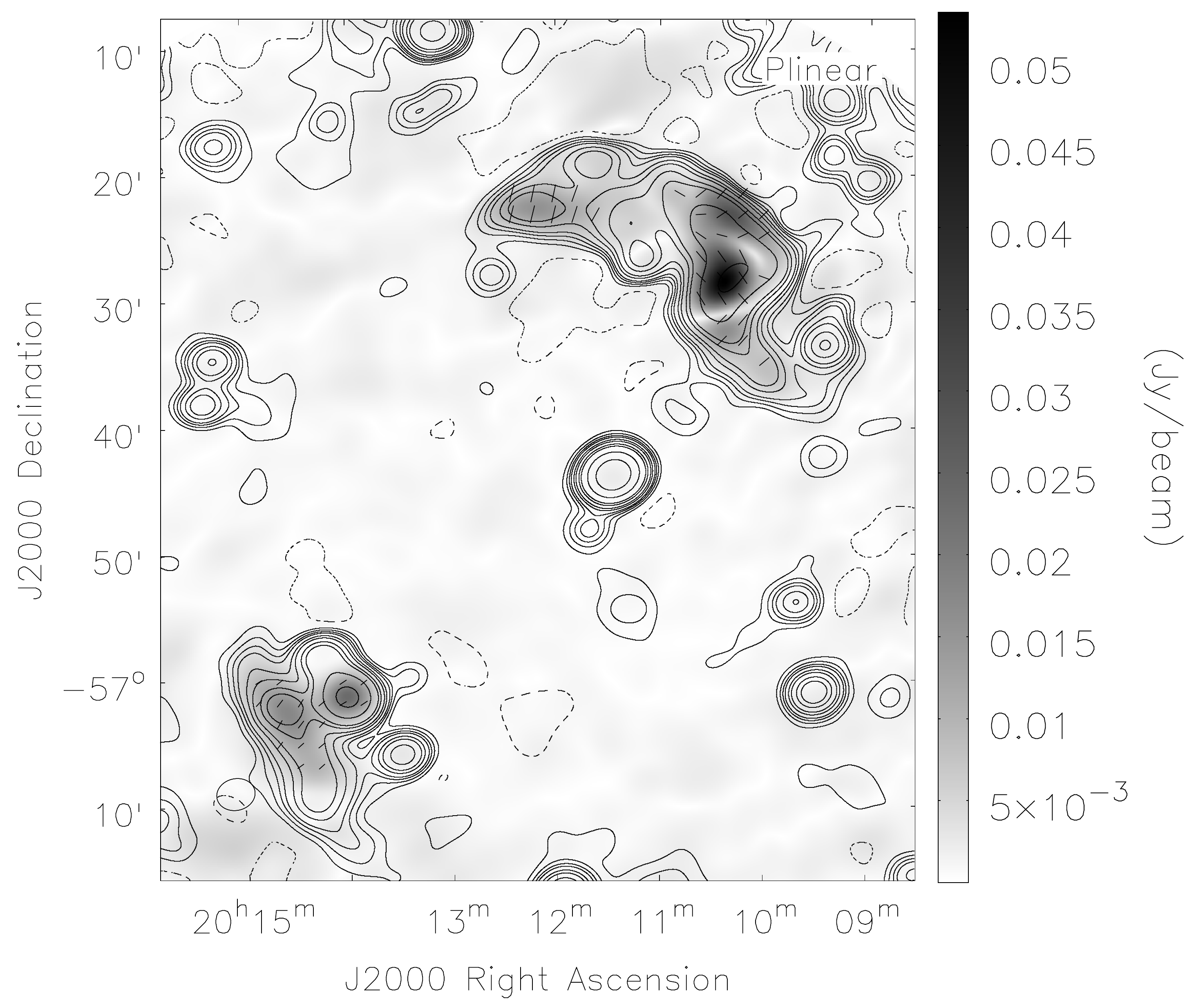}
		\label{fig:1826_poli}
	\end{subfigure}
	\begin{subfigure}[b]{0.7\textwidth}
		{\includegraphics[width=\textwidth]{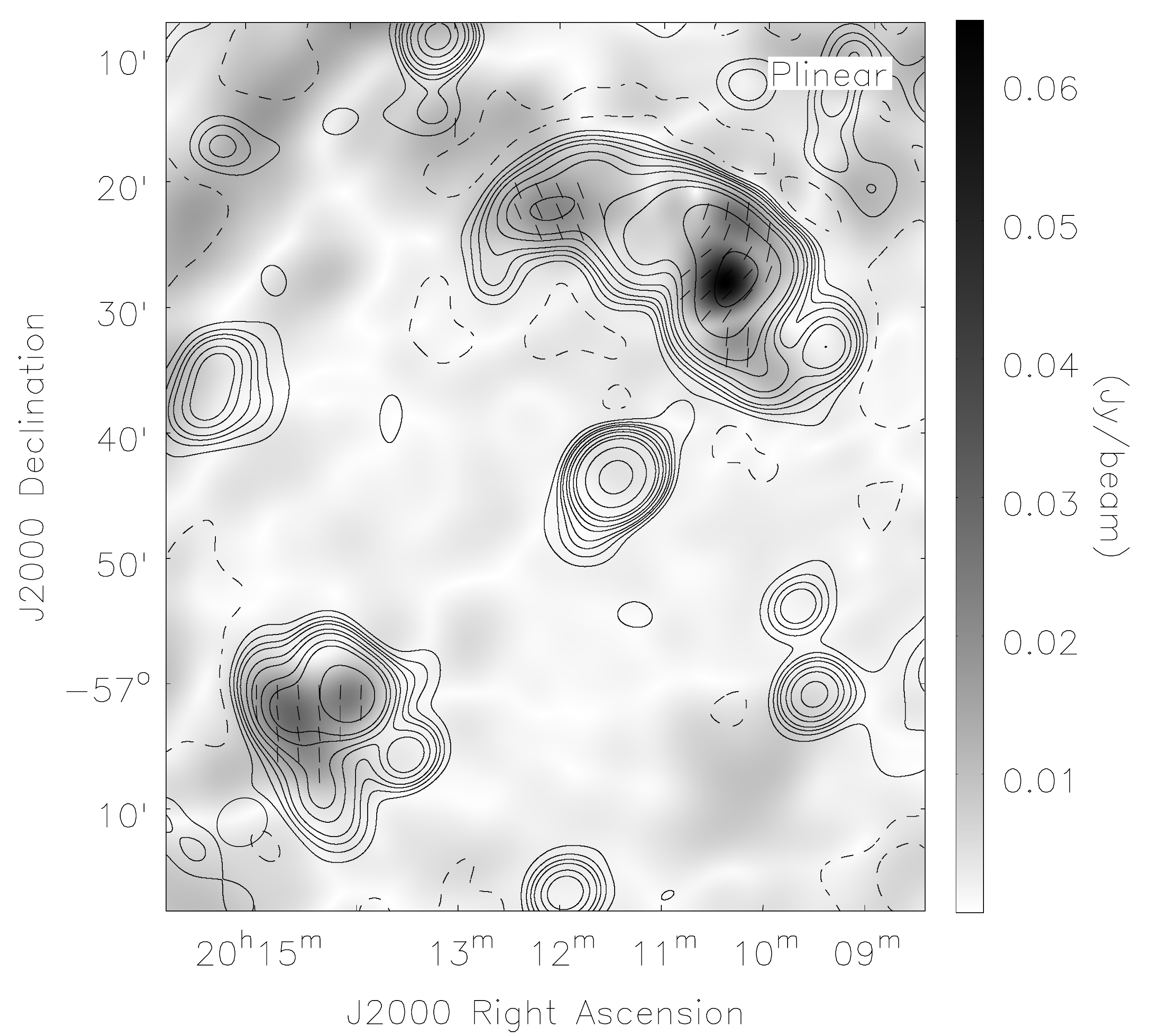}}
		\label{fig:1372_poli}
	\end{subfigure}
	\caption{Linearly-polarized maps of A3667 at 1826 (\textsc{top}) and 1372 (\textsc{bottom}) MHz. Contours are Stokes I at the corresponding frequency, as per Figures \ref{fig:1826_StokesI} and \ref{fig:1372_StokesI}. Vectors are the polarization angle rotated by 90$\degree$ to represent the direction of the magnetic field, derived where the polarized flux density exceeds 10 $\sigma_{\rm{rms}}$ and $\sigma_{\rm{rms}}= 0.73$ (1.25) mJy beam$^{-1}$ at 1826 (1372) MHz.\label{fig:linpolmaps}}
\end{figure*}

We present linearly-polarized images at 1826 and 1372 MHz in Figure \ref{fig:linpolmaps}. Contours are Stokes I at the corresponding frequency, and vectors display the polarization angle of emission from the relics for regions where the flux density exceeds a 10$\sigma_{\rm{rms}}(P)$ threshold, where $\sigma_{\rm{rms}}(P)=0.73\,(1.25)$ mJy beam$^{-1}$ at 1826 (1372) MHz.

The polarisation vectors have been both rotated by 90\degree\, to represent the direction of the magnetic field and corrected for Galactic Faraday rotation using the map of \cite{2012A&A...542A..93O}. We estimate the Galactic Faraday depth contribution in the direction of the NW (SE) relic to be $5.29\pm15.10$ ($3.33\pm9.014$) rad m$^{-2}$; we correct by subtracting a rotation of 8.17\degree\,(5.14\degree) at 1826 MHz and 14.47\degree\,(9.11\degree) at 1372 MHz.

\begin{table*}

\caption{Source population of the A3667 field as observed with KAT-7. Corresponding integrated flux densities at 843 MHz have been retrieved from the SUMSS catalogue \citep{2003MNRAS.342.1117M} for each source. RA and DEC values quoted from the fit to the 1826 MHz map. Flux densities were only fitted to sources where the primary beam response is $>50$ per cent and source flux density rises to S$_{\rm{peak}} \geq7\sigma_{\rm{rms}}$.\\ \textsc{upper panel}: Field sources. \textsc{lower panel}: Sources present within extended emission from relics. \label{sourcepop}}

\begin{threeparttable}
\centering
\begin{tabular}{l | cc  | r  r | r | rc p{2cm}}
\hline\hline
 & & & \multicolumn{3}{ c |}{Integrated Flux Density} & & \\
 \hline
 & & & \multicolumn{2}{ c |}{KAT-7} & SUMSS & &  \\
SUMSS Catalog Name & RA & DEC & 1826 MHz & 1372 MHz & 843 MHz & $\alpha_{843}^{1826}$ & C13 ID\tnote\dag & Notes \\
 & (J2000) & (J2000) & (mJy) & (mJy) & (mJy) & & & \\
\hline
SUMSS J200930--570058 & 20h09m30s & $-57\degree00^{\prime}52^{\prime\prime}$ & 82.4$\pm$4.8 & 93.2$\pm$6.3 & 127.2$\pm$4.0 & $-0.56\pm0.09$ & - & - \\
SUMSS J200940--565354 & 20h09m40s & $-56\degree53^{\prime}52^{\prime\prime}$ & 38.3$\pm$3.1 & 37.3$\pm$3.1 & 50.2$\pm$1.8 & $-0.35\pm0.11$\tnote\ddag & 13 & - \\
SUMSS J201127-564358 & 20h11m27s & $-56\degree43^{\prime}44^{\prime\prime}$ & 471.2$\pm$23.7 & 589.1$\pm$29.9 & 1030.8$\pm28.3$\tnote{$\ast$} & $-1.01\pm0.07$ & 1 & B2007-569 \\
SUMSS J201155--571655 & 20h11m56s & $-57\degree16^{\prime}45^{\prime\prime}$ & 80.9$\pm$5.3 & 76.4$\pm$5.9 & 66.3$\pm$2.2 & $0.25\pm0.10$ & - & - \\
SUMSS J201309--560842 & 20h13m10s & $-56\degree08^{\prime}33^{\prime\prime}$ & -\tnote{$\star$} & 129.0$\pm$8.3 & 142.8$\pm$4.5 & $-0.13\pm0.09$ & - & Index is $\alpha_{843}^{1372}$ \\
SUMSS J201518--563440 & 20h15m16s & $-56\degree34^{\prime}31^{\prime\prime}$ & 49.7$\pm$3.4 & \multirow{2}{*}{149.6$\pm$10.5}\tnote{$\diamond$} & 81.8$\pm$2.7 & $-0.64\pm0.10$ & - & - \\
SUMSS J201524--563805 & 20h15m21s & $-56\degree37^{\prime}42^{\prime\prime}$ & 38.3$\pm$4.2 & & 58.9$\pm$2.0 & $-0.55\pm0.15$ & - & - \\
\hline
SUMSS J200925--563327 & 20h09m25s & $-56\degree33^{\prime}27^{\prime\prime}$ & 54.8$\pm7.2$ & 74.5\tnote{$\circ$} & 108.0$\pm3.5$ & $-0.88\pm0.18$& 5 & Within NW relic. \\
SUMSS J201330--570552 & 20h13m32s & $-57\degree05^{\prime}51^{\prime\prime}$ & 50.3$\pm4.5$ & 73.9\tnote{$\circ$} & 122.1$\pm$3.8 & $-1.15\pm0.12$ & - & Within SE relic. \\
\hline\hline
\end{tabular}

\begin{tablenotes}
	\item[\dag] This column lists source ID numbers from Table 1 of \cite{2013MNRAS.430.1414C} where catalogued. 
	\item[\ddag] This source appears embedded in a region of extended emission at 1826 MHz (see Figure \ref{fig:1826_StokesI}) so this spectral index should be considered an upper limit.
	\item[$\star$] The SUMSS catalog lists two components for B2007-569; we quote the sum of the flux densities from both and derive the integrated spectrum from this value. 
	\item[$\ast$] Primary beam response $<$ 50 per cent.
	\item[$\diamond$] Due to the lower resolution at 1372 MHz, these sources cannot be resolved separately.
	\item[$\circ$] Fits could not converge, so we present peak flux densities read directly from the image.
\end{tablenotes}
\end{threeparttable}

\end{table*}

\section{Analysis}\label{sec:AN}
\subsection{Source Identification}
In these data, we have shown strong detections of both large-scale diffuse emission and compact sources at both frequencies. The two relics exhibit similar morphologies to those presented in \cite{1997MNRAS.290..577R} and \cite{2003PhDT.........3J}. Additionally, we detect many of the brighter sources seen in other work, as revealed by (for example) comparison with the Sydney University Molonglo Sky Survey (SUMSS; \citealt{1999AJ....117.1578B}) catalogue. With the resolution available to KAT-7, we have angular scale sensitivity that is intermediate to that of \cite{1997MNRAS.290..577R} and \cite{2013MNRAS.430.1414C}.

The source populations of Figures \ref{fig:1826_StokesI} and \ref{fig:1372_StokesI} were catalogued and had positions and flux densities extracted using the \textsc{casa} task \texttt{imfit}\footnote{http://casaguides.nrao.edu/index.php?title=Imfit}. A catalogue of sources common to both figures is presented in Table \ref{sourcepop}, containing integrated flux densities from both KAT-7 maps as well as the integrated flux density of the corresponding source in the SUMSS catalogue \citep{2003MNRAS.342.1117M}. The spectral index was fitted using the values at 1826 and 843 MHz; values derived are consistent with non-thermal extragalactic source populations. 

Integrated flux densities for each relic were computed using the \textsc{fitflux} flux density fitting routine \citep{2007BASI...35...77G} deriving values of $1.92\pm0.14$ ($0.34\pm0.02$) Jy for the NW (SE) relic. At 1372 MHz, the NW (SE) relic has an integrated flux density of $2.46\pm0.17$ ($0.35\pm0.02$) Jy -- see Table \ref{fluxcompfloat} for a summary. All integrated flux measurements were conducted after subtracting contributions from point sources identified by \cite{2013MNRAS.430.1414C} -- see \S\ref{sec:sourcesub} for details on the subtraction process.

With the resolution of KAT-7, the compact sources and extended relic in the SE form one structure. The compact sources are PMN J2014-5701 (a head-tail radio source) and SUMSS J201330-570552 (a suspected FRII; \citealt{2003PhDT.........3J}). With the resolution of ATCA however, previous work (\citealt{2003PhDT.........3J}, 2004) has resolved the separate features in the SE relic. Hence for the KAT-7 integrated flux density, the measurements were restricted to the elongated outer part of the structure. 

The integrated flux density of the NW relic is consistent with the work of \cite{1997MNRAS.290..577R} where an integrated flux of 2.4 Jy at 1.4 GHz was derived, although there is discrepancy with the integrated flux density of 3.7 Jy derived by \cite{2003PhDT.........3J}. The negative bowl around the NW relic in Figure \ref{fig:1372_StokesI} suggests some missing flux on the largest angular scales, which may partially account for this discrepancy. For the SE relic, the integrated flux densities of $0.34\pm0.02$ ($0.35\pm0.02$) Jy at 1826 (1372) MHz agree reasonably well with the integrated flux density of 0.30$\pm$0.02 Jy from \cite{2003PhDT.........3J} at 1.4 GHz.

\subsection{Polarized Structure}\label{sec:polstruc}
As can be seen in Figure \ref{fig:linpolmaps}, we detect linearly-polarized emission from both relics at each frequency. There is little ambient emission detected elsewhere at 1826 MHz, whereas at 1372 MHz there is a greater level of polarized emission outside the relics. We attribute this higher level of ambient emission to off-axis leakage effects. The linear polarization vectors in Figure \ref{fig:linpolmaps} have been rotated by 90$\degree$ to represent the direction of the magnetic field. 

From Figure \ref{fig:linpolmaps}, at 1372 MHz the magnetic field vectors lie approximately along the long axis of the SE relic, a result consistent with DSA theory and observations (e.g. \citealt{2010Sci...330..347V}; \citealt{2012A&A...546A.124V}). For the NW relic however the picture is harder to interpret as a result of the contorted morphology. Nevertheless, some departure from the expected alignment is observed, which may indicate the merger axis is slightly tilted, similar to the case of PSZ1 G096.89+24.17 \citep{2014arXiv1408.2677D}. In such a case, the ICM would produce an observable rotation in the polarization angle measured in the more distant relic, but would not be seen in the closer relic. This deviation is not inconsistent with optical observations that suggest a merger event close to the plane of the sky (Johnston-Hollitt et. al 2008; Owers et al. 2009)
 
At 1826 MHz however, the magnetic field vectors lie approximately perpendicular to the long axis of the relics, suggesting they are experiencing Faraday rotation. Given the relatively small frequency difference between the datasets, an rotation of approximately 60$\degree$ in magnetic field direction might imply a high rotation measure (RM). However, although there is low fractional bandwidth difference between the two maps, in wavelength-squared-space the difference is more apparent. The magnetic field vector rotation of $60\degree$ implies a RM of $-158.3\pm7.5$ rad m$^{-2}$, accounting for the $n\pi$ ambiguity in RM measurements (see for example \citealt{2009IAUS..259..591H}). Such a value is consistent with previous RM estimates for the NW relic \citep{2004rcfg.proc...51J}.

When deriving polarization fractions for the relics, \textsc{fitflux} was used to determine the integrated polarized flux. At 1826 MHz we recover an integrated flux density of $216\pm13$ ($53\pm8$) mJy for the NW (SE) relic; corresponding to a polarization fraction of $11.45\pm1.07$ ($15.6\pm2.2$) per cent. At 1372 MHz we recover an integrated flux density of $163\pm11$ ($55\pm7$) mJy for the NW (SE) relic; a polarization fraction of $6.6\pm0.6$ ($15.6\pm2.6$) per cent. For comparison, \cite{2013MNRAS.430.1414C} derived fractional polarization values of 8.5 (14) per cent for the NW (SE) relic at 2.3 GHz. However, they note that -- as a result of the resolution of the single-dish Parkes radio telescope -- the NW relic proved difficult to constrain, as the polarized emission blends smoothly with the weaker background emission. \cite{2003PhDT.........3J} measured a polarization fraction of 10--40 per cent for the NW relic at 1.4 GHz with ATCA. 

Additionally, we detect no polarized emission from the bridge region above the noise level. This is consistent with the observations of \cite{2013MNRAS.430.1414C} and the suggestion that turbulence generated in the wake of the passing merger shock would disrupt the ordered structures and large-scale alignment necessary for polarized emission on these scales.

\begin{table}
\caption{Integrated flux density of the total intensity and linearly-polarized intensity of diffuse radio emission from the relics of Abell 3667 at 1826 and 1372 MHz. Also listed are the corresponding sensitivity and resolution at each frequency. Note that we measure the Stokes I integrated flux density after performing point source subtraction as described in \S\ref{sec:sourcesub}.\label{fluxcompfloat}}
\centering
\begin{tabular}{l | l | r | r }
\hline\hline
				       & $\nu_{\rm{obs}}$ (MHz)    & 1372                     & 1826 \\
\hline
\multirow{3}{*}{NW relic} & S$_{\rm{int}}$(I) (Jy)   & $2.47\pm0.17$     & $1.92\pm0.14$ \\
				     & S$_{\rm{int}}$(P) (Jy)  & $0.16\pm0.01$  & $0.22\pm0.01$  \\
	   			    & P/I (per cent)		  & $6.63\pm0.57$ & $11.45\pm1.07$ \\
\hline 						            
\multirow{3}{*}{SE relic} & S$_{\rm{int}}$(I) (Jy)   & $0.35\pm0.02$     & $0.34\pm0.02$ \\
				     & S$_{\rm{int}}$(P) (Jy)  & $0.06\pm0.01$  &  $0.05\pm0.01$ \\
	   			    & P/I (per cent)		  & $15.60\pm2.60$ & $15.60\pm2.20$ \\
\hline
				& $\theta$ (arcsec)						& 251 & 180 \\
				& $\sigma_{\rm{rms}}$(I) (mJy beam$^{-1}$) 	& 2.18 & 1.30 \\
\hline\hline
\end{tabular}
\end{table}

\subsection{Spectral Index Profile}
From Figures \ref{fig:1826_StokesI} and \ref{fig:1372_StokesI} is is possible to map the spectral index variation between 1826 and 1372 MHz. To improve the fractional bandwidth over which the spectrum is mapped, however, we employ ancillary data from SUMSS to map the spectral index between 1826 and 843 MHz. We initially convolve the SUMSS map to match the resolution of KAT-7 at 1826 MHz ($184\times150$ arcsec) before masking emission from the relics at 10 mJy beam$^{-1}$ ($\sim7\sigma_{\rm{rms}}$) to mitigate contributions from noise.

\begin{figure*}
	\begin{subfigure}[tl]{0.47\textwidth}
		\includegraphics[width=\textwidth]{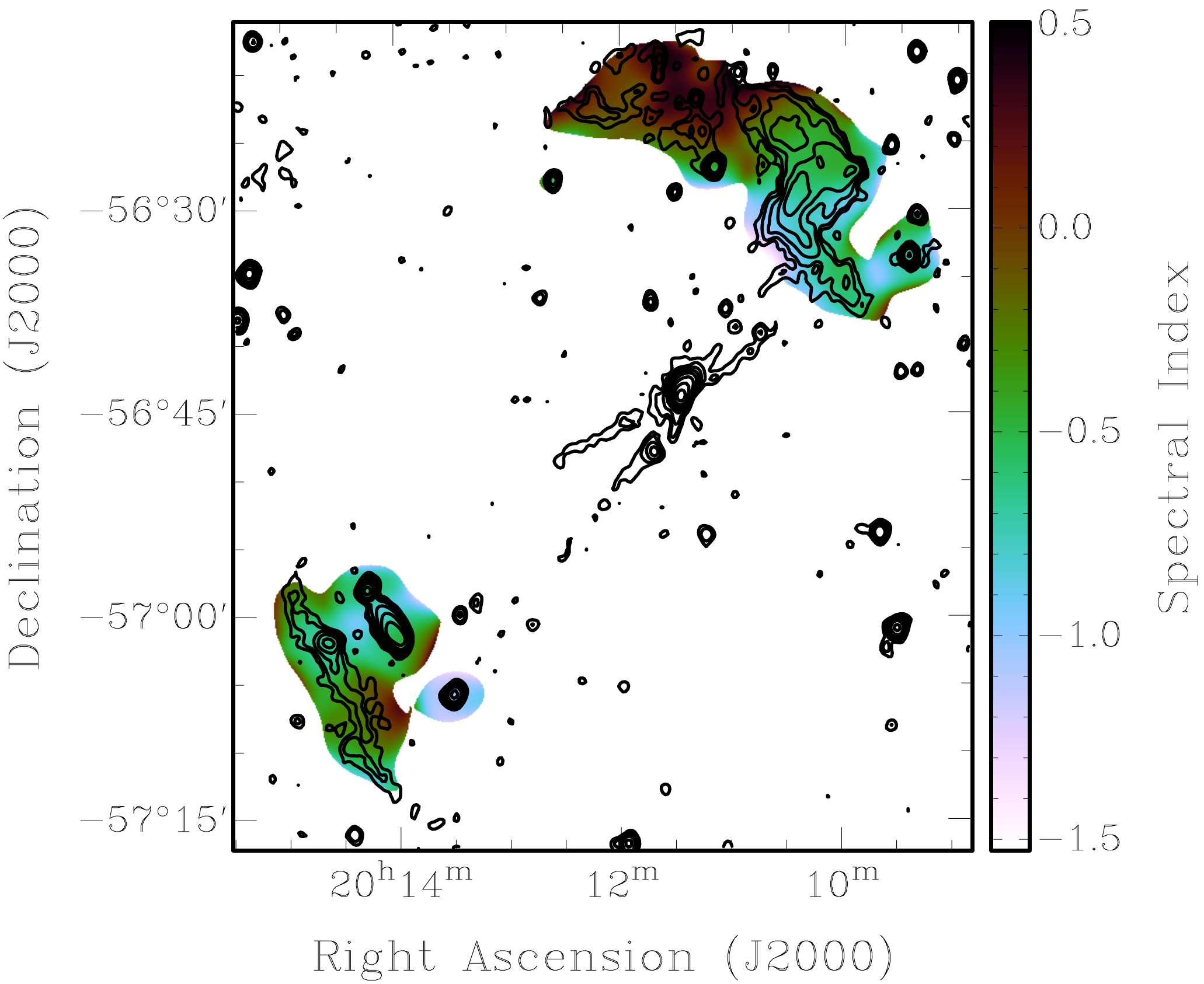}
		\label{fig:spix_full}
	\end{subfigure}
	\qquad
	\begin{subfigure}[tr]{0.46\textwidth}
		{\includegraphics[width=\textwidth]{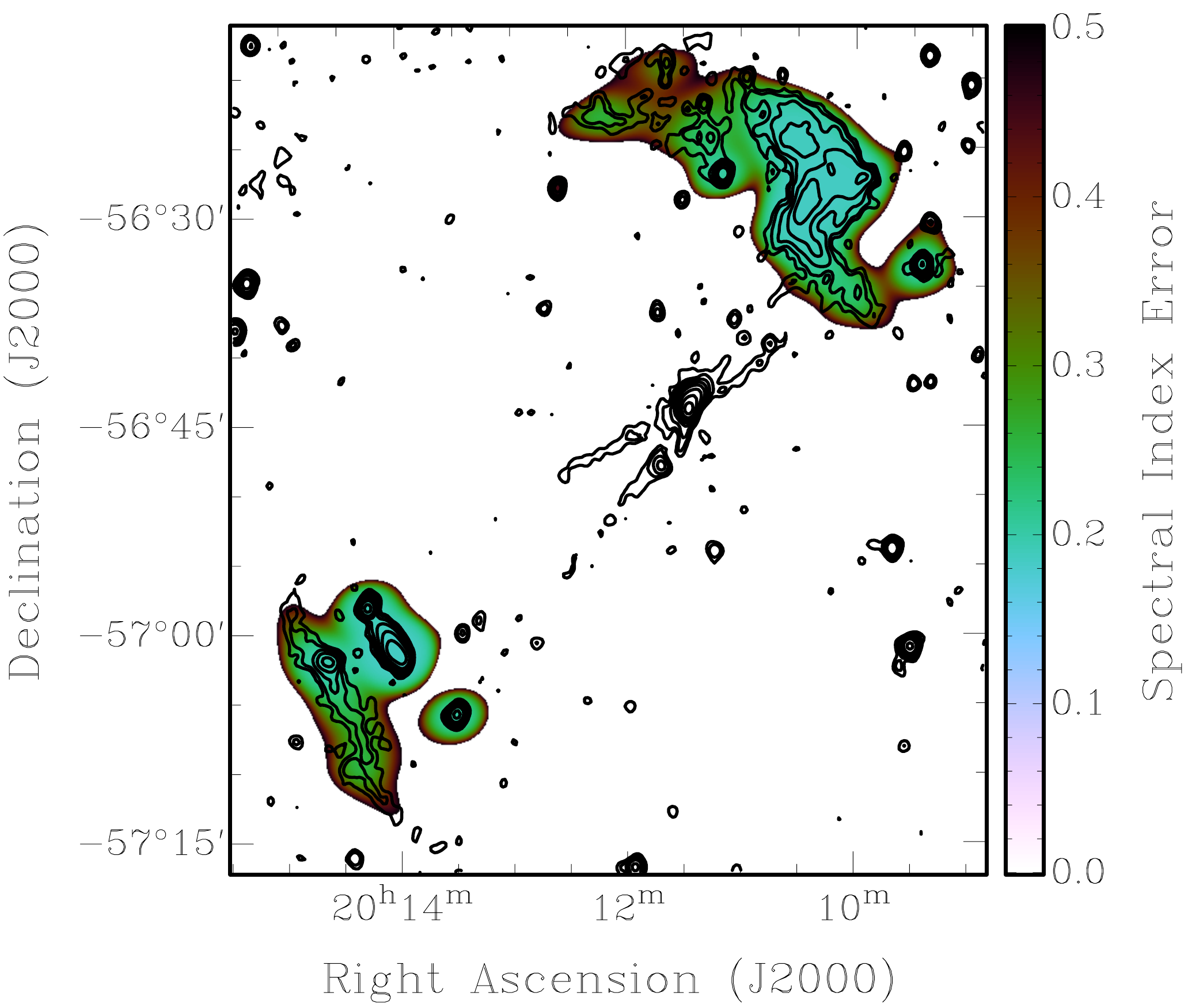}}
		\label{fig:spix_err}
	\end{subfigure}
	
	\caption{Spectral index profile (\textsc{left}) and associated uncertainty (\textsc{right}) of the radio relics of A3667 between 843 and 1826 MHz (36 and 16cm) at a resolution of $184\times150$ arcsec (resolution of KAT-7 at 1826 MHz). Contours are full-resolution Stokes I at 843 MHz, from SUMSS, starting at 3 mJy beam$^{-1}$ and scaling by a factor of 2. (A colour version of this Figure is available in the online journal). \label{fig:spix}}
\end{figure*}

The resultant spectral index map (along with the associated uncertainty) is shown in Figure \ref{fig:spix}. For the NW relic, the spectral index is typically in the range $-1.0\pm0.2$ to $-0.5\pm0.2$, marginally flatter than the profile derived by \cite{1997MNRAS.290..577R} where $-1.5 < \alpha_{\rm{1.4 GHz}}^{\rm{2.4 GHz}} < -1.0$, which may indicate curvature in the spectrum. The spectrum steepens to $-1.2\pm0.3$ toward the cluster centre, flattening to approximately $-0.3\pm0.2$ toward the leading edge of the relic; such variation may indicate differences in the ages of the electrons responsible for the radio emission.

While the spectrum inverts for the Eastern spur of the NW relic, this is likely due to differences in sensitivity to faint diffuse emission resulting from resolution differences between SUMSS ($53\times45$ arcsec with sensitivity 1 mJy beam$^{-1}$) and KAT-7 ($184\times150$ arcsec with sensitivity 1.30 mJy beam$^{-1}$). The resulting difference in sensitivity to large-scale diffuse emission is visible in the breaking up of the SUMSS contours in this region (see Figure \ref{fig:spix}). 

The SE relic exhibits a flatter spectrum, typically in the range $-0.7\pm0.2$ to $-0.4\pm0.2$. This is consistent with DSA being the acceleration mechanism (the flattest spectrum possible according to DSA is $-0.5$) and in agreement with earlier evidence of DSA from the alignment of the magnetic field vectors with the long axis of the relic (see Figure \ref{fig:linpolmaps} and \S\ref{sec:polstruc}). Both spectral index ranges are broadly consistent with the spectral index variation derived by \cite{2003PhDT.........3J} although with our moderate-resolution data we do not resolve the spectral index structure of the SE relic. Given that MOST data at 843 MHz are common to both this work and \cite{2003PhDT.........3J} such consistency is not unexpected. 

\cite{2014MNRAS.445..330H} find integrated spectral index values of $\alpha_{120}^{1400} = -0.9\pm0.1$ for both the NW and SE relics. From our KAT-7 map, the spectral index of the NW is typically $\alpha_{843}^{1826} = -0.8\pm0.2$, consistent with the work of Hindson et al. Additionally, Hindson et al. find the 843 MHz flux of the NW relic inconsistent with their 4-point spectral index map; their interpretation is that MOST may be missing some flux on scales of order 30 arcmin, comparable to the size of the NW relic. Such missing flux may be responsible for the spectral index shape recovered previously. With KAT-7, we are sensitive to structures up to 22 arcmin in size; we cannot discount the possibility that some flux is missing on the largest angular scales. For the SE relic, we find a typical spectral index of $-0.5\pm0.2$, inconsistent with the integrated spectrum derived by Hindson et al.

\subsection{Equipartition Magnetic Field}
The magnetic field strength of A3667 can be estimated using the equipartition field strength equation from \cite{2005AN....326..414B}. The equipartition magnetic field strength is given by
\begin{equation}\label{eqn:beq}
\centering
B_{\rm{eq}} =  \left\{ \frac{4 \rm{\pi} \hspace{0.05cm} (2\alpha + 1) \hspace{0.05cm} (K_0 + 1) \hspace{0.05cm} \mathit{ I_{\nu} } \hspace{0.05cm} \mathit{E}_p^{1-2\alpha} \hspace{0.05cm} (\nu/2c_1)^{\alpha}}{(2\alpha -1) \hspace{0.05cm} c_2(\alpha) \hspace{0.05cm} c_4(i) \hspace{0.05cm} \ell } \right\}^{1/(\alpha+3)}
\end{equation}
where $\alpha$ is the spectral index, $I_{\nu}$ is the synchrotron intensity at frequency $\nu$; $E_{\rm{p}}$ is the proton rest energy and $K_0$ is the ratio of the number densities of protons and electrons (Beck \& Krause 2005). $c_1$, $c_2$ and $c_4$ are constants given by the following:
\begin{subequations}
\begin{align}
 c_1 = & \frac{3\rm{e}}{4\rm{\pi} \rm{m_e}^3 \rm{c}^5} = 6.26428\times10^{18} \text{erg}^{-2}\text{ s}^{-1}\text{ G}^{-1}\\
 c_2 = & \frac{1}{4} c_3 \frac{(\gamma_e + 7/3)}{(\gamma_e +1)}\Gamma[(3\gamma_e-1)/12]  \times \Gamma[(3\gamma_e+7)/12] \label{eq:c2}\\
 c_3 = & \sqrt{3}\rm{e}^3/(4\rm{\pi} \rm{m_e} \rm{c}^2) = 1.86558\times10^{-23} \text{erg} \text{ G}^{-1} \text{ sterad}^{-1} \\
 c_4 = & [\cos(i)]^{(\gamma_{\rm{e}}+1)/2}
\end{align}
\end{subequations}
where $\gamma_e$ is the spectral index of the electron energy spectrum, related to the spectral index by $\alpha = (\gamma_e-1)/2$; $i$ is the inclination angle with respect to the sky plane and is assumed to be constant for $c_4$ to be valid as presented \citep{2005AN....326..414B}; m$_{\rm{e}}$ is the mass of the electron and c is the speed of light. A value of $i = 0$ denotes a face-on view, which is a reasonable approximation for A3667 as the merger event is thought to be occurring approximately in the plane of the sky (for example \citealt{2010ApJ...715.1143F}). Typical ratios of proton-to-electron number density are $K_0 = 100-300$ \citep{2004MNRAS.352...76P}; we take a mid-range value of $K_0 = 200$ for our derivation. $\ell$ is the line-of-sight path length, which must be assumed for the calculation; we assume a cylindrical geometry, measuring a path length of 720 arcsec for the NW relic and 350 arcsec for the SE relic. Substituting $\gamma_e = 2\alpha +1$ in Equation \ref{eq:c2}, we derive:
\begin{equation*}
c_2 = \frac{1}{4} c_3 \frac{(2\alpha + 10/3)}{(2\alpha + 2)} \hspace{0.1cm} \Gamma[(6\alpha+2)/12] \hspace{0.1cm} \Gamma[(6\alpha+10)/12] 
\end{equation*}

\begin{figure}
	\centering
	\includegraphics[width=1.0\columnwidth]{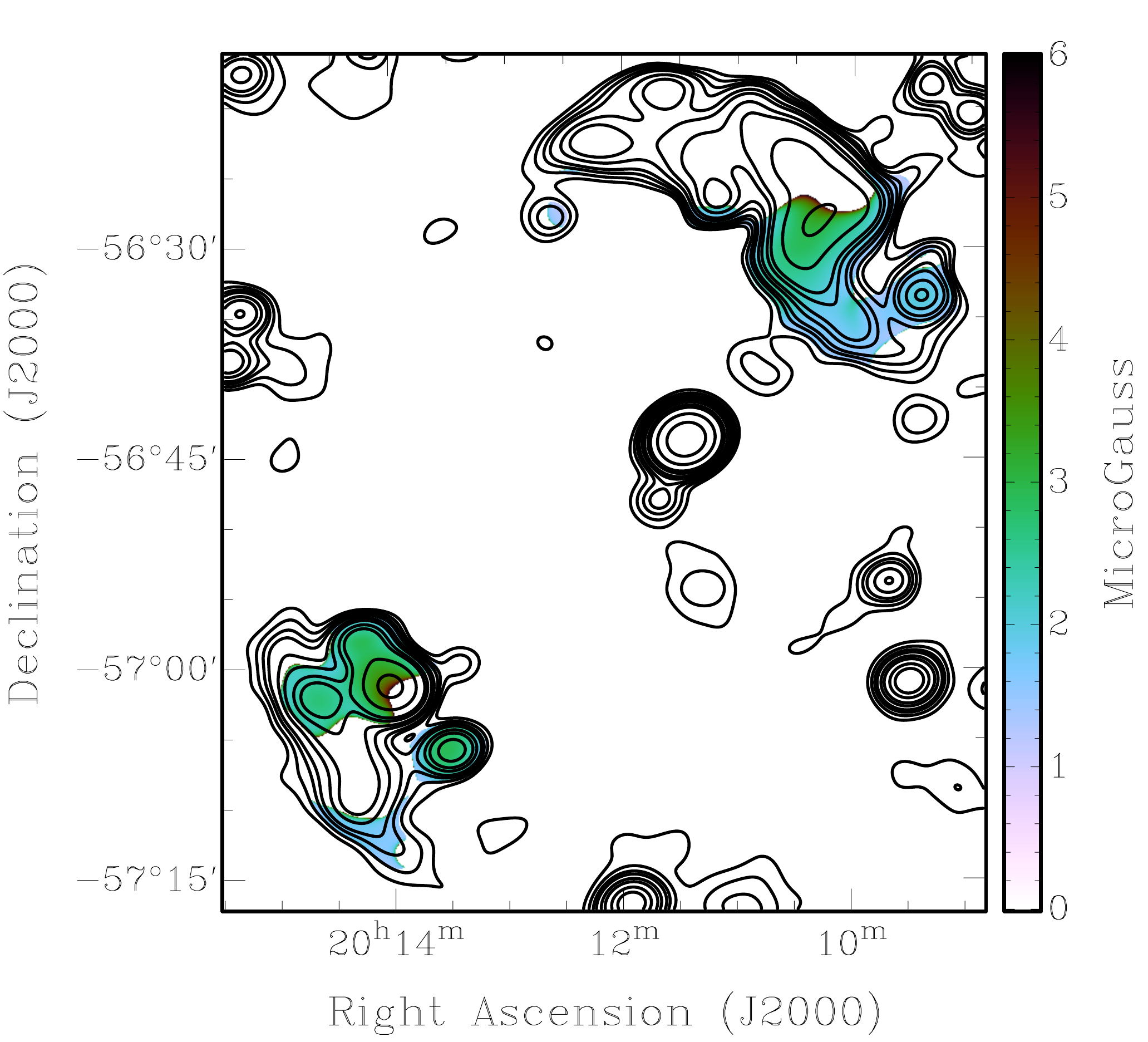}
	\caption{Estimated equipartition magnetic field strength in the radio relics of A3667. Contours are KAT-7 Stokes I at 1826 MHz as per Figure \ref{fig:1826_StokesI}. Estimates for the SE relic are largely contaminated by the background FRII. (A colour version of this Figure is available in the online journal).}
	\label{fig:beq}
\end{figure}

The resultant map of equipartition magnetic field strength ($B_{\rm{eq}}$) is presented in Figure \ref{fig:beq}. Large regions have no $B_{\rm{eq}}$ estimate due to the fact that Equation \ref{eqn:beq} diverges for $\alpha > -0.5$. For the SE relic, the regions where estimates exist are largely contaminated by flux from either of the secondary sources\footnote{These sources are SUMSS J201330-570552; a suspected FRII source \citep{2003PhDT.........3J} and the head-tail radio source PMN J2014-5701.} which have previously been resolved separately (e.g. Johnston-Hollitt 2003, 2004). Therefore we restrict our discussion of the equipartition magnetic field strength to the NW relic.

For the NW relic, the flat spectrum of the Eastern spur of this relic prohibits estimation of $B_{\rm{eq}}$. However, the estimates for the remainder of the relic yield values of $B_{\rm{eq}}$ in the range $1.5 - 3.0 \, \, \mu\rm{G}$; consistent with equipartition estimates from previous work (e.g. \citealt{2004rcfg.proc...51J}; see Table \ref{bfields}).

\subsection{Radio Bridge}\label{sec:sourcesub}
With the modest resolution of KAT-7, we are able to probe the diffuse low surface-brightness emission from A3667 on scales up to 22 (30) arcminutes at 1826 (1372) MHz. Given the extent suggested for the proposed radio bridge (20 arcmin; \citealt{2013MNRAS.430.1414C}) KAT-7 should possess sufficient sensitivity to recover this feature. However, as seen in Figure \ref{fig:1826_StokesI}, only patchy low-level flux is recovered between the central galaxy and the NW relic; this emission is coincident with a number of point sources that appear in the SUMSS catalogue \citep{2003MNRAS.342.1117M}. 

Extrapolating from the surface brightness of the bridge as presented in \cite{2013MNRAS.430.1414C} of 0.21 -- 0.25 $\,\, \rm{\mu}$Jy arcsec$^{-2}$ at 2.3 GHz, using the spectral index of $\alpha = -1.21$ derived therein, we calculate 1826 MHz (1372 MHz) surface brightness in the range 0.29 -- 0.33 (0.39 -- 0.45) $\,\, \rm{\mu}$Jy arcsec$^{-2}$. From these surface brightness values, we should expect to recover flux density of the order of 12 -- 14 (35 -- 40) mJy beam$^{-1}$ with KAT-7, corresponding to the order of 10 (17)$\times \sigma_{\rm{rms}}$ -- a level not recovered in Figures \ref{fig:1826_StokesI} and \ref{fig:1372_StokesI}. At low frequency (226 MHz) \cite{2014MNRAS.445..330H} recover a surface brightness from the bridge region that is below the significance expected from extrapolation of the surface brightness as presented by \cite{2013MNRAS.430.1414C}. 

In order to further investigate the nature of the radio `bridge', we extrapolate the flux densities of the 13 sources above 3\,mJy identified by Carretti et. al. (2013; Table 1) to the observing frequencies covered in this work and subtract these sources from the data. All sources are modelled as a simple power-law fit, with the exception of head-tail radio galaxy B2007-569, which requires more careful modelling.

\begin{table}
\begin{center}
\caption{Integrated flux density measurements for B2007-569 from the literature and derived in this work, with their respective flux scales and the scale factor required to convert to the scale of Baars et al. (1977). Integrated flux density is quoted before any scaling factor is applied.\label{tab:flux_models}}
\begin{threeparttable}
\begin{tabular}{llrcc}
\hline\hline
Freq. & Telescope                          & Flux Density    & Flux Scale & Scale Factor \\
(MHz) &                                          & (mJy)               &                   &                      \\
\hline
120 & MWA\tnote\dag				& 7300$\pm400$ & BT & -- \\
149 & MWA\tnote\dag				& 6500$\pm400$ & BT & -- \\
180 & MWA\tnote\dag				& 5100$\pm300$ & BT & -- \\
226 & MWA\tnote\dag				& 4200$\pm200$ & BT & -- \\
408 & MOST\tnote{$\star$}                 & 1920$\pm150$ & W69 & 0.97 \\
843 & MOST\tnote{$\star$}              & 1031$\pm28$ & (c)B77 & n/a \\
1372 & KAT-7\tnote{$\ast$}         & 589$\pm30$ & B77 & n/a \\
1383 & ATCA\tnote{$\ast$}	    & 497$\pm25$ & B77 & n/a \\
1588 & ATCA\tnote{$\ast$}	    & 416$\pm21$ & B77 & n/a \\
1793 & ATCA\tnote{$\ast$}	    & 365$\pm18$ & B77 & n/a \\
1826 & KAT-7\tnote{$\ast$}         & 471$\pm24$ & B77 & n/a \\
2202 & ATCA\tnote{$\ast$}	    & 304$\pm15$ & B77 & n/a \\
2300 & ATCA\tnote{$\diamond$}            & 280$\pm14$\tnote{$\circ$} & B77 & n/a \\
2407 & ATCA\tnote{$\ast$}	    & 284$\pm14$ & B77 & n/a \\
2817 & ATCA\tnote{$\ast$}	    & 246$\pm12$ & B77 & n/a \\
3022 & ATCA\tnote{$\ast$}	    & 229$\pm12$ & B77 & n/a \\
3300 & ATCA\tnote{$\diamond$}            & 219$\pm11$\tnote{$\circ$} & B77 & n/a \\
4850 & Parkes\tnote{$PMN$} & 173$\pm12$ & (c)B77 & n/a \\
\hline
\multicolumn{5}{l}{Flux scale notation:} \\
\multicolumn{5}{l}{BT: Bootstrap method using known sources in the field -- see }\\
\multicolumn{5}{l}{\hspace{0.5 cm} \protect{\cite{2014MNRAS.445..330H}} for details.} \\
\multicolumn{5}{l}{W69: \protect{\cite{1969MNRAS.142..229W}}} \\
\multicolumn{5}{l}{B77: \protect{\cite{1977A&A....61...99B}}} \\
\multicolumn{5}{l}{(c) implies the flux scale is consistent with the given scale.} \\
\hline\hline
\end{tabular}

\begin{tablenotes}
	\item[\dag] \cite{2014MNRAS.445..330H}
	\item[\ddag] Molonglo Reference Catalog; \cite{1981MNRAS.194..693L}
	\item[$\star$] SUMSS; \cite{2003MNRAS.342.1117M} 
	\item[$\ast$] This work
	\item[$\diamond$]  \cite{2013MNRAS.430.1414C}
	\item[$PMN$] Parkes-MIT-NRAO (PMN) Survey; \cite{1994ApJS...91..111W}
	\item[$\circ$] These error values taken as 5 percent of the integrated flux density.
\end{tablenotes}
\end{threeparttable}

\end{center}
\end{table}

Table \ref{tab:flux_models} presents the integrated flux density of B2007-569 between 408 and 4850 MHz, both derived in this work and sourced from the literature. The flux densities have been measured using a number of different instruments (listed in Table \ref{tab:flux_models}) sensitive to emission on several different scales. The ATCA flux densities derived from our analysis of the archival CABB data were obtained by running \textsc{fitflux} on images made in each 200\,MHz subband. We include the measured MWA flux densities from \cite{2014MNRAS.445..330H} in Table \ref{tab:flux_models} although we omit them from the fitting as uncertainties remain due to inconsistencies between the analytic beam model \citep{2012ApJ...755...47W} and beam response as revealed during commissioning \citep{2014arXiv1410.0790H}.

With the exception of the Molonglo Reference Catalog (MRC; \citealt{1981MNRAS.194..693L}) flux density, all values in Table \ref{tab:flux_models} were derived in work where the flux scale was either set according to the \cite{1977A&A....61...99B} scale or shown to be consistent with it. The MRC employed the \cite{1969MNRAS.142..229W} flux scale. \cite{1977A&A....61...99B} note that the \cite{1969MNRAS.142..229W} scale overestimates the flux density by 3 per cent compared to the \cite{1977A&A....61...99B} scale; hence this can be accounted for by a scaling factor 0.97.

The SUMSS flux scale was set through both measurements at 843 MHz and interpolation between the MRC at 408 MHz and the Parkes catalog at 2700 MHz \citep{1999AJ....117.1578B} following the method of \cite{1991AuJPh..44..743H}. This flux scale is primarily tied to the historically-determined flux density of PKS B1934-638 (among others) using the \cite{reynolds94} scale, which broadly agrees with the \cite{1977A&A....61...99B} scale \citep{reynolds94}. Hence no scaling factor is necessary for the SUMSS flux densities used in this work. The PMN catalog flux density values have been shown to be consistent with the \cite{1977A&A....61...99B} scale \citep{1994ApJS...91..111W}.

\begin{figure}
	\centering
	\begin{subfigure}[t]{0.9\columnwidth}
		\includegraphics[width=\columnwidth]{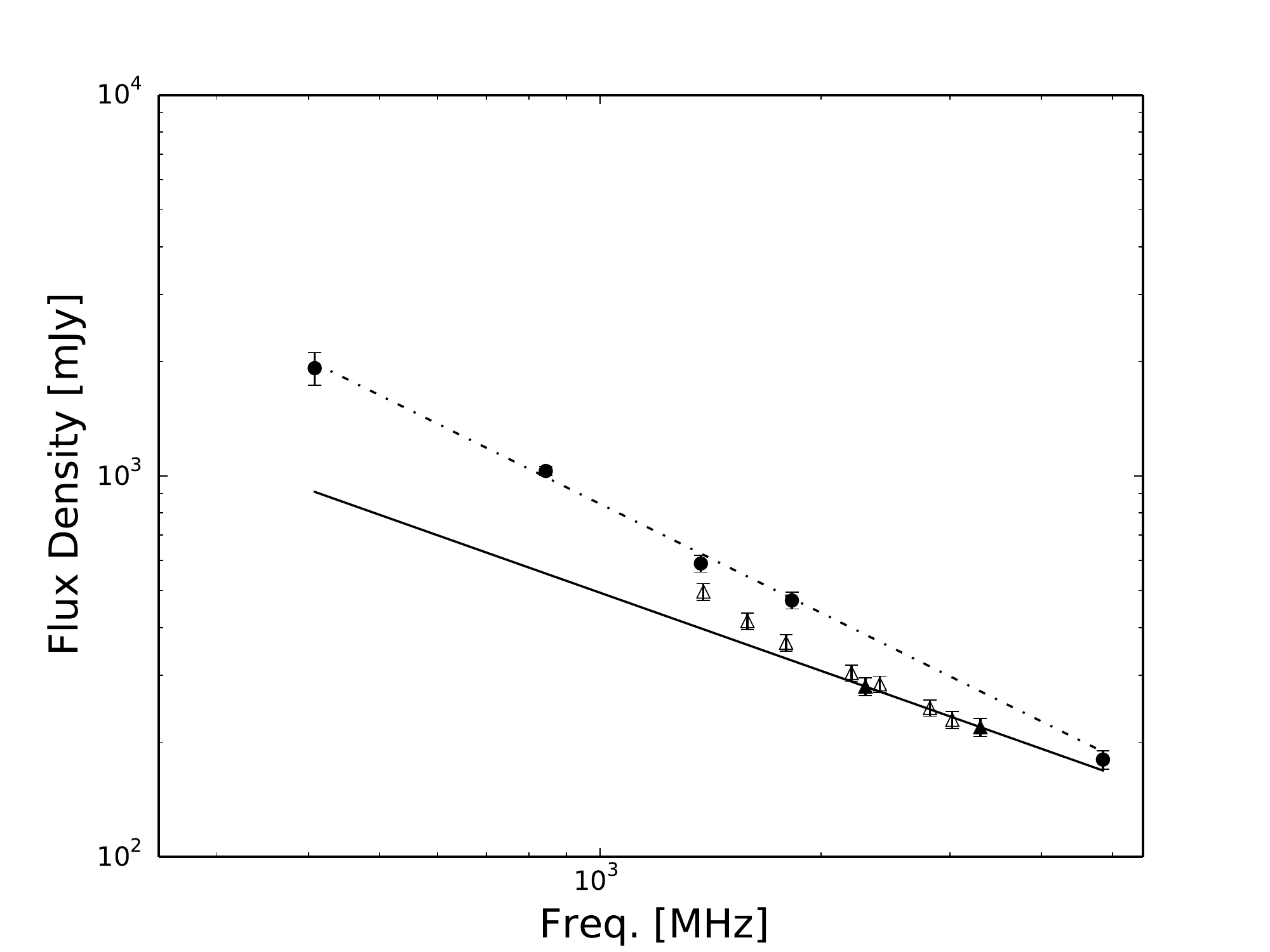}
	\end{subfigure}
	\begin{subfigure}[b]{0.9\columnwidth}
		\includegraphics[width=\columnwidth]{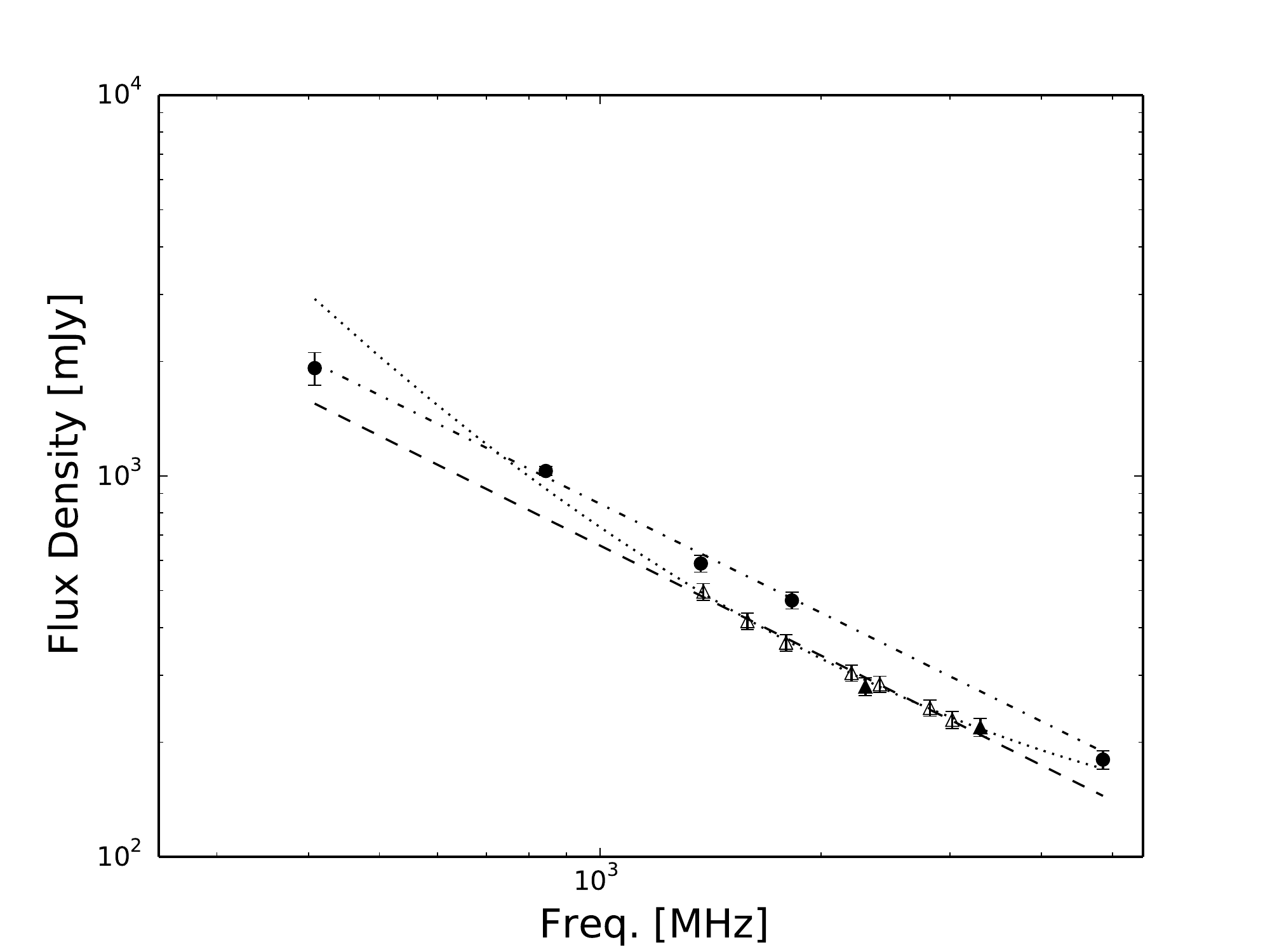}
	\end{subfigure}
	\caption{Integrated spectrum of B2007-569 between 408 MHz and 4850 MHz. Both panels present identical data with different fits shown. Top panel presents the power-law fits to the MOST/KAT-7 data between 408--1826\,MHz (dot-dashed, $\alpha=-0.949\pm0.104$) and ATCA data of Carretti et al. (2013; solid line, $\alpha = -0.681 \pm0.489$). Lower panel presents fits to the ATCA data as analysed in this work. Dashed slope (dotted line) is a power-law (second-order polynomial) fit to the ATCA data (hollow triangles) analysed in this work, with the power-law fit yielding a spectral index $\alpha = -0.958\pm0.162$. Also presented is the fit in the range 408--1826\,MHz for reference.}
	\label{fig:b2007spectrum}
\end{figure}

We present the scaled flux densities in Figure \ref{fig:b2007spectrum}. From inspection, the integrated flux density of B2007-569 appears to follow a power-law between 408 and 1826 MHz; we derive a spectral index of $\alpha=-0.949\pm0.104$. While the integrated flux densities derived from our analysis of the ATCA data agree well with those of \cite{2013MNRAS.430.1414C} the flux densities derived from the MOST and KAT-7 data are in excess of the flux densities derived using the ATCA.

The flux density of B2007-569 as measured with ATCA in this work is well-described by a power-law fit with spectral index $\alpha=-0.958\pm0.162$. We also fit a polynomial of the form $\rm{log}(S) = \emph{A} + \emph{B}\,\rm{log}\nu + \emph{C}\,(\rm{log}\nu)^2$ from which we derive a spectral index of the form $\alpha = -4.752 + 1.144 (\rm{log \nu})$ where $\nu$ is measured in MHz. \cite{2014MNRAS.445..330H} find a spectral index fit of $\alpha=-1.1\pm0.1$ between 120 and 4850 MHz, consistent with the spectral index values derived for B2007-569 from this work.

\begin{figure*}
	\begin{subfigure}[l]{0.45\textwidth}
		\includegraphics[width=\textwidth]{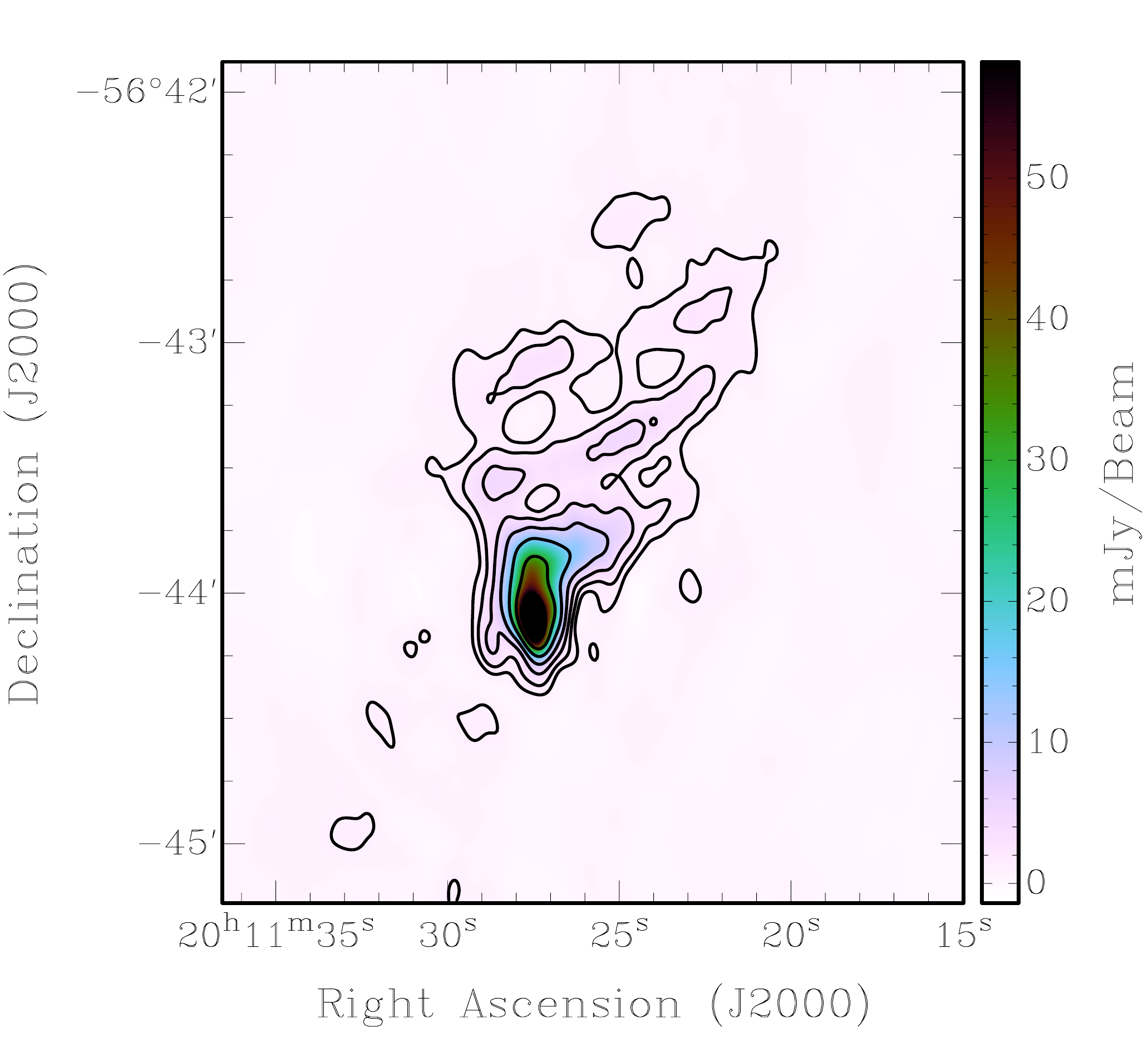}
		\caption{Stokes I}
		\label{fig:b2007_SUMSS}
	\end{subfigure}
	\quad
	\begin{subfigure}[r]{0.45\textwidth}
		{\includegraphics[width=\textwidth]{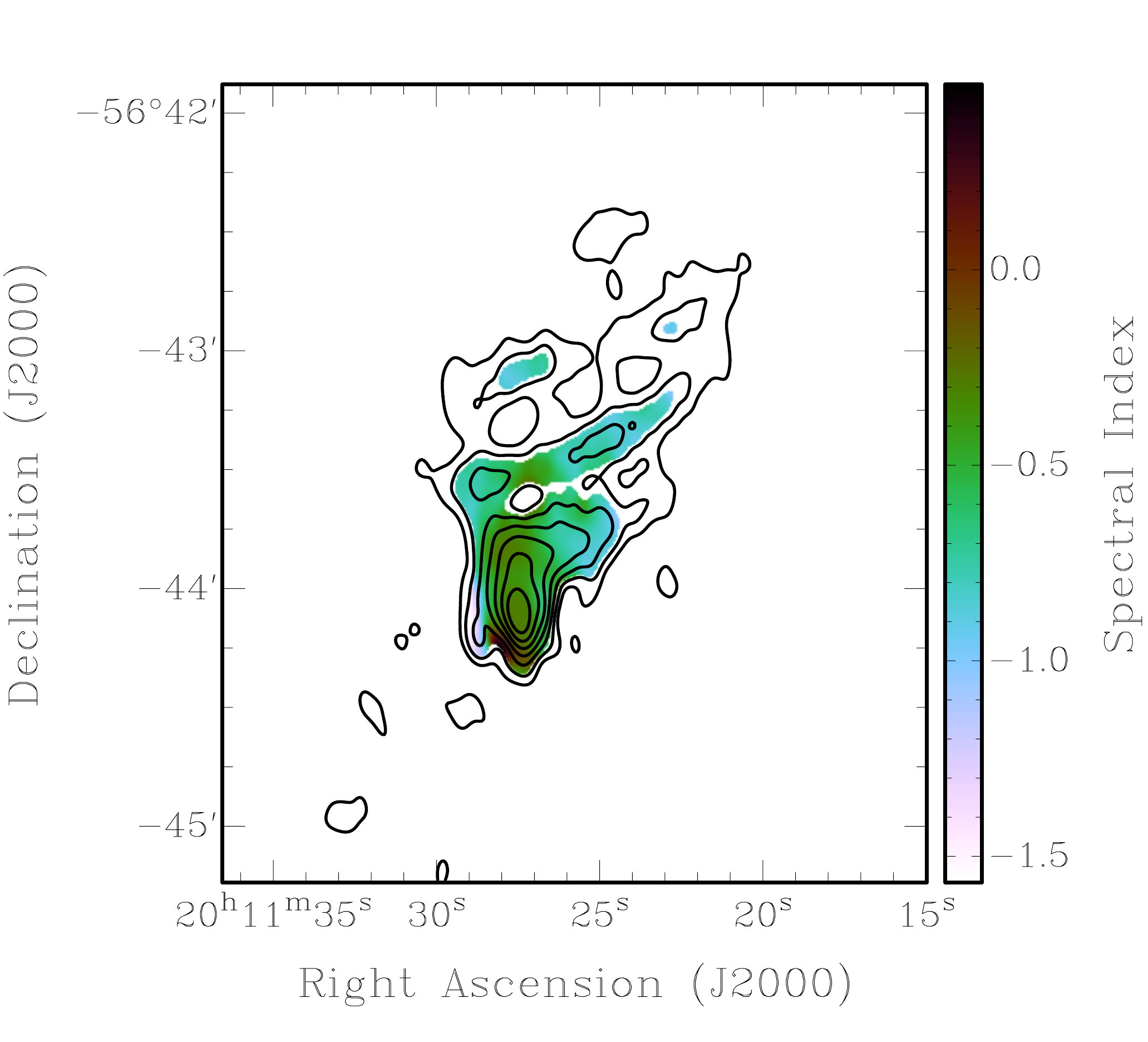}}
		\caption{Spectral Index}
		\label{fig:fig:b2007_mod}
	\end{subfigure}
	\quad
	\begin{subfigure}[c]{0.45\textwidth}
		{\includegraphics[width=\textwidth]{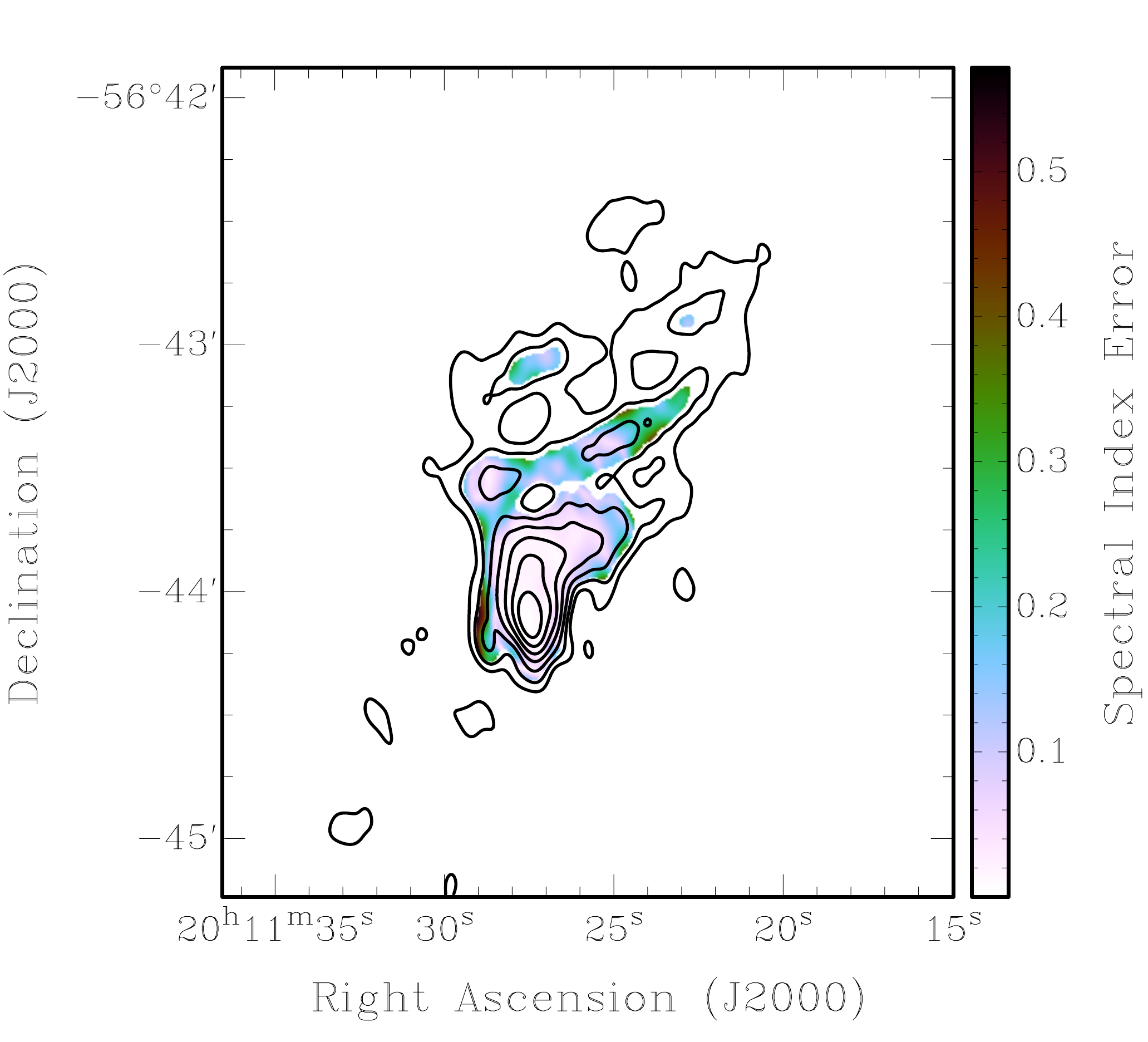}}
		\caption{Spectral Index Error}
		\label{fig:b2007_residuals}
	\end{subfigure}
\caption{ATCA maps of B2007-569 produced using MFS-clean on the CABB band in the frequency range 1288--3124 MHz. \textsc{top left}: Stokes I image at a reference frequency of 1400 MHz. \textsc{top right}: Spectral index. \textsc{lower}: Spectral index error. Contours are the Stokes I ATCA data starting at 1 mJy/beam and scaling by a factor 2. Beam size is $9\times5$ arcsec. $\sigma_{\rm{rms}}=0.2$ mJy beam$^{-1}$. (A colour version of this Figure is available in the online journal).}
\label{fig:b2007_modelling}
\end{figure*}

B2007-569 is not point-like at the resolution of the instruments considered here. We use the archival ATCA data to obtain a higher-resolution clean-component model (CCM) for subtraction. The wide bandwidth available with CABB yields suitable frequency coverage for multi-frequency synthesis (MFS; \citealt{2011A&A...532A..71R}) \texttt{clean} in \textsc{casa}. To obtain the CCM, we performed \texttt{mfs}-\texttt{clean} using two Taylor terms, with the reference frequency set to the 1400\,MHz. Following \texttt{mfs}-\texttt{clean}, a spectral index map and its associated error map are produced; these are presented in Figure \ref{fig:b2007_modelling} alongside the Stokes I map at 1400\,MHz.

From Figure \ref{fig:b2007_modelling}, B2007-569 has a bright central component with the extended tails fading in surface-brightness. The spectral index map indicates that the spectrum of the central component is relatively flat (typically $\alpha=-0.5$), with a marked steepening toward the fainter tails. Both of these results are consistent with previously published images of B2007-569 (for example \citealt{1997MNRAS.290..577R}). Imaging B2007-569 across the full CABB bandwidth has indicated that the tail components rapidly fade below the detectable flux limit, yielding images morphologically consistent with those of \cite{2013MNRAS.430.1414C}.

In the upper band, only the point sources were subtracted, in order to investigate their contribution to the flux from the bridge region. In the lower band, with the improved sensitivity to larger angular scales, B2007-569 was subtracted as well as the point sources. The frequency-dependent ATCA CCM was scaled to account for the position of B2007-569 relative to the phase centre of each KAT-7 pointing and subtracted from the \emph{uv} data.

\begin{figure*}
	\begin{subfigure}[tl]{0.45\textwidth}
		\includegraphics[width=\textwidth]{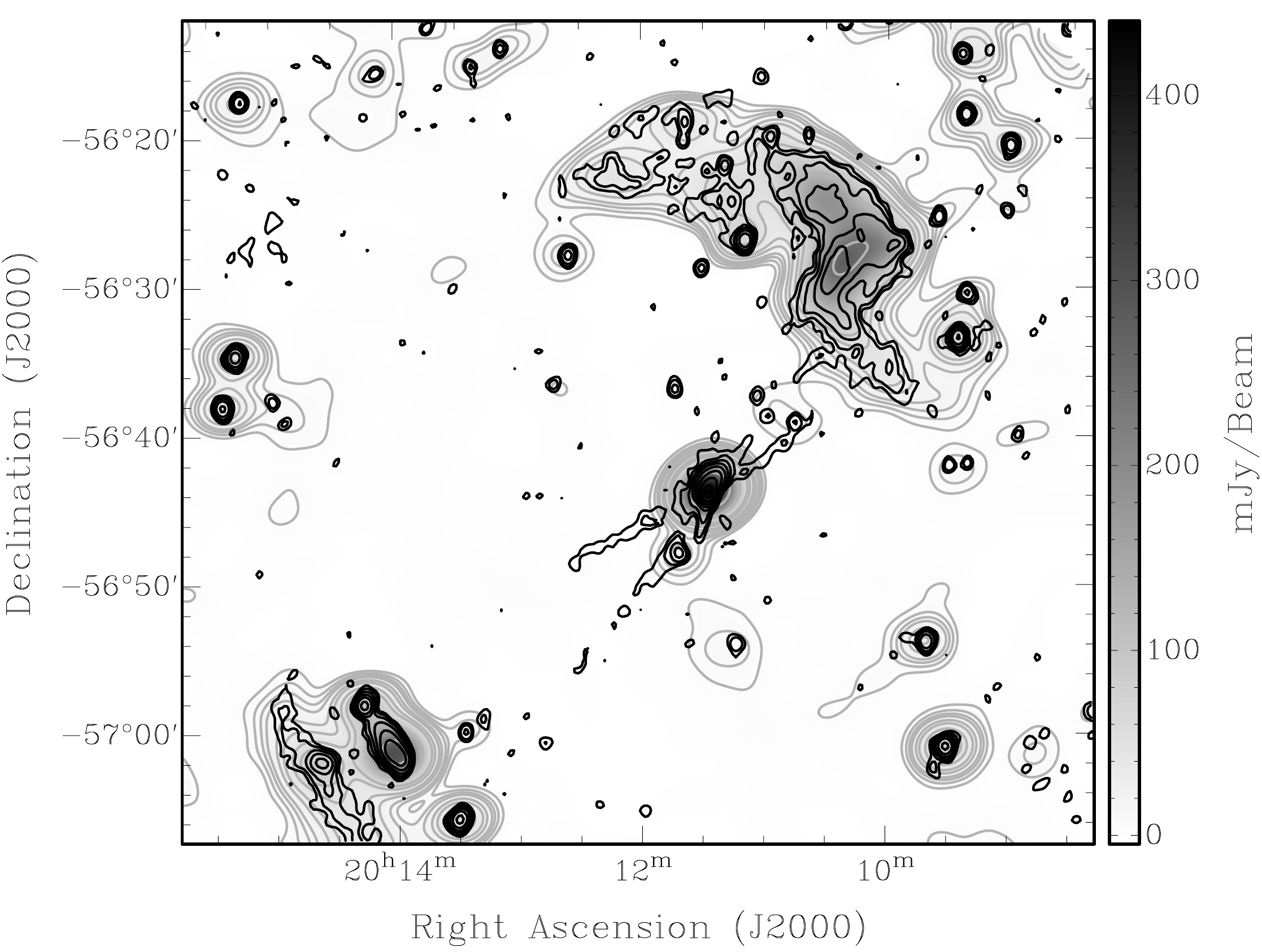}
		\caption{Full-resolution image of A3667 at 1826 MHz, prior to source subtraction. $\sigma_{\rm{rms}}=1.30$ mJy beam$^{-1}$.\vspace{0.3cm}}
		\label{fig:a3667_1826SUMSS}
	\end{subfigure}
	\quad
	\begin{subfigure}[tr]{0.45\textwidth}
		{\includegraphics[width=\textwidth]{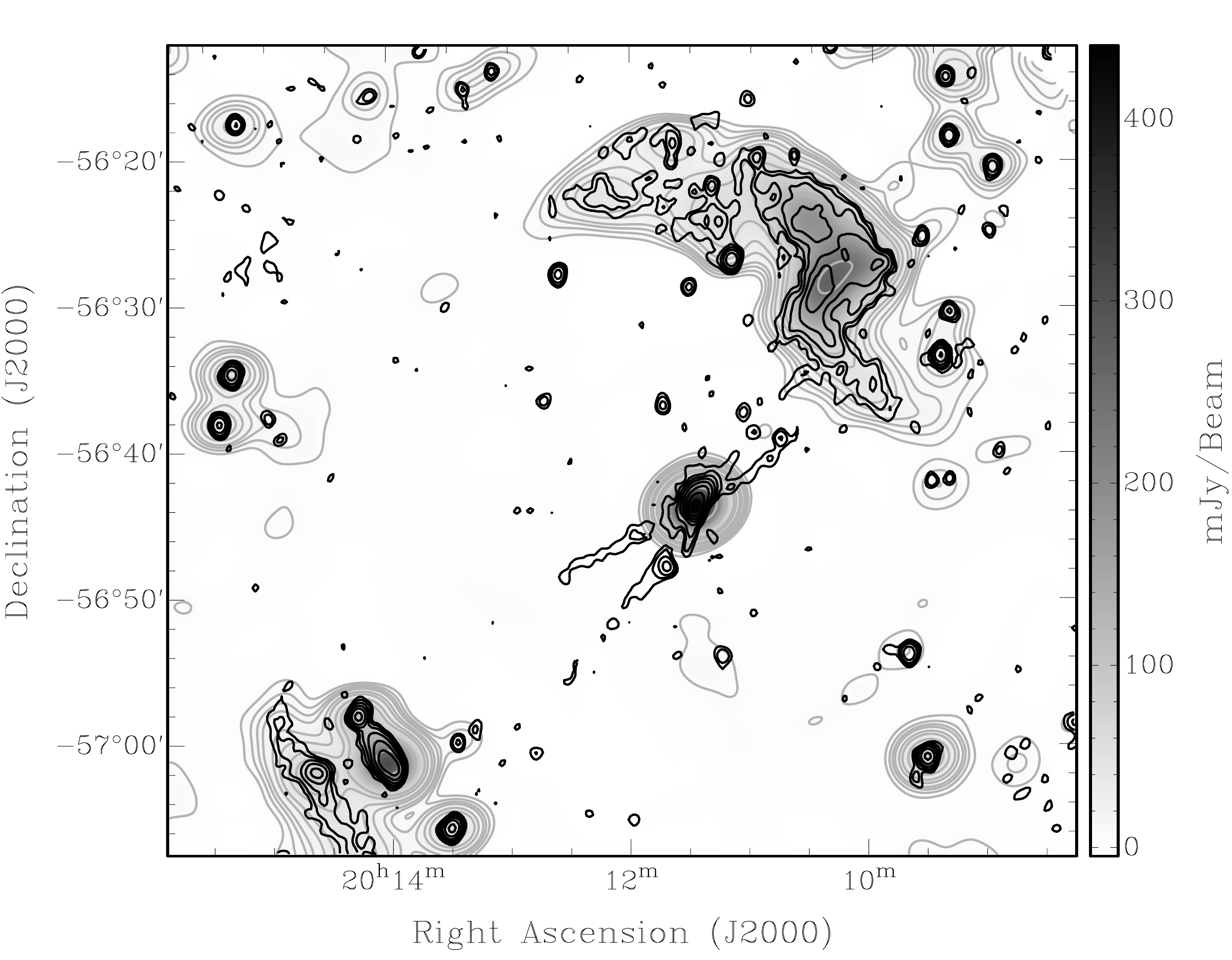}}
		\caption{Full-resolution image of A3667 at 1826 MHz, with point sources (as identified in \citealt{2013MNRAS.430.1414C}) subtracted. $\sigma_{\rm{rms}}=1.30$ mJy beam$^{-1}$. \vspace{0.3cm}}
		\label{fig:1826_uvsub}
	\end{subfigure}
	\begin{subfigure}[bl]{0.45\textwidth}
		{\includegraphics[width=\textwidth]{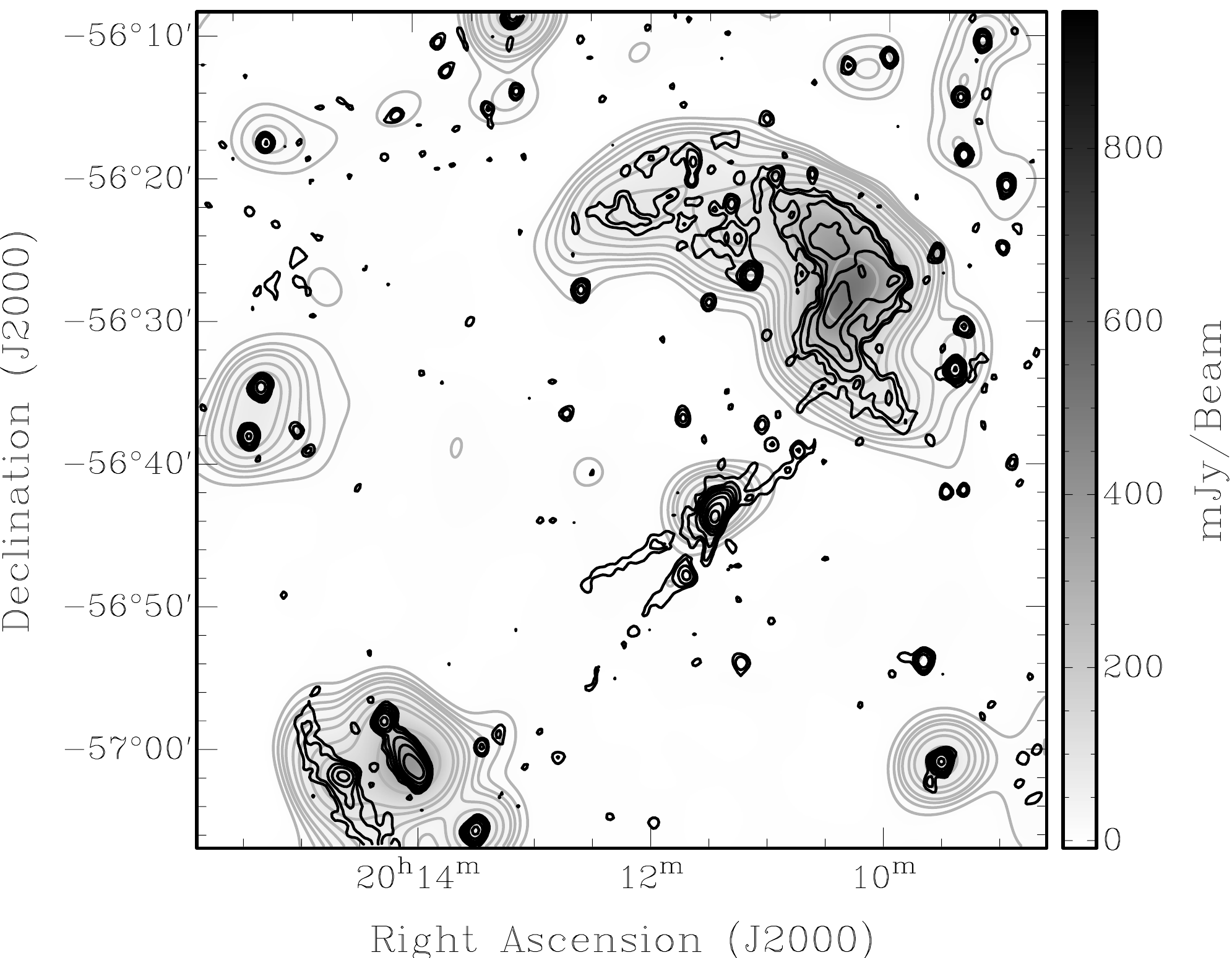}}
		\caption{Full-resolution image of A3667 at 1372 MHz, with head-tail radio galaxy B2007-569 and point sources (as identified in \cite{2013MNRAS.430.1414C} subtracted. $\sigma_{\rm{rms}}=2.18$ mJy beam$^{-1}$.}
		\label{fig:a3667_1372sub}
	\end{subfigure}
	\quad
	\begin{subfigure}[br]{0.45\textwidth}
		{\includegraphics[width=\textwidth]{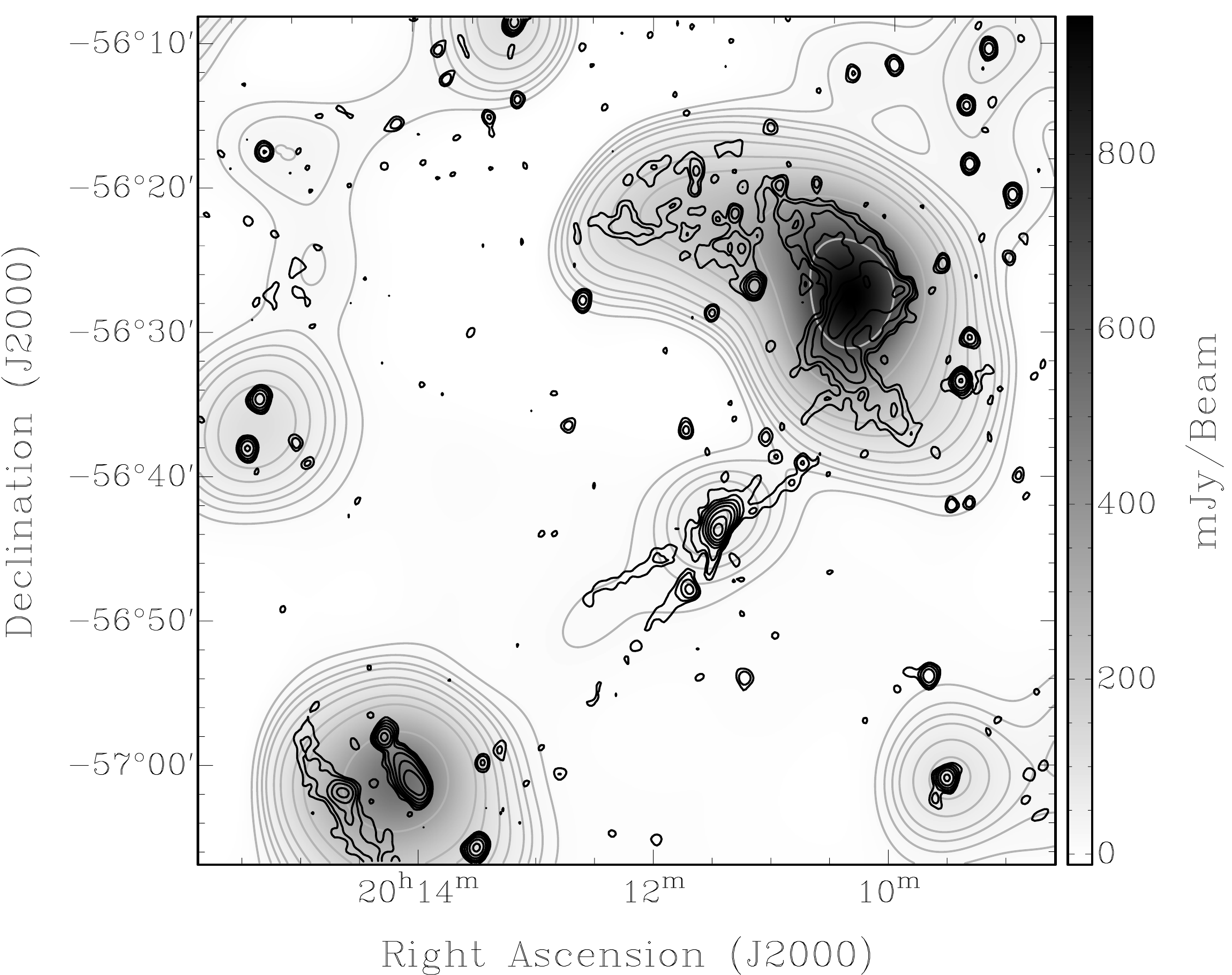}}
		\caption{\emph{uv}-tapered image A3667 at 1372 MHz,  with resolution $7.18^{\prime}\times6.50^{\prime}$, to match the resolution of the the higher-frequency observations in \cite{2013MNRAS.430.1414C}. $\sigma_{\rm{rms}}=3.55$ mJy beam$^{-1}$.}
		\label{fig:a3667_1372sub_uvtaper}
	\end{subfigure}
\caption{Stokes I maps of A3667 observed with KAT-7. Grey contours are positive Stokes I as observed with KAT-7 at [1,2,3,4,6,8,12,16,32]$\times 3\sigma_{\rm{rms}}$, where $\sigma_{\rm{rms}}$ is stated in the image caption. Black contours are SUMSS data at 843 MHz, starting at 3 mJy beam$^{-1}$ and scaling by a factor 2. Note that the X-shaped emission extending from the location of B2007-569 has been identified as imaging artefact. Maps at matching frequencies (i.e. Figures \ref{fig:a3667_1826SUMSS} and \ref{fig:1826_uvsub}; Figures \ref{fig:a3667_1372sub} and \ref{fig:a3667_1372sub_uvtaper}) are set to matching colour scales to facilitate comparison.}
\label{fig:a3667_uvsubmaps}
\end{figure*}

The results of subtraction are presented in Figure \ref{fig:a3667_uvsubmaps}. From the upper band maps -- Figures \ref{fig:a3667_1826SUMSS} and \ref{fig:1826_uvsub} -- the point sources in the bridge region appear to dominate the contribution to the detected flux, as minimal residual emission remains above the noise level. In the lower band maps, subtraction of the ATCA model of B2007-569 results in strong residual emission co-located with the head-tail radio galaxy, shown in Figure \ref{fig:a3667_1372sub}. We also apply a \emph{uv}-taper to our 1372 MHz map to match the resolution of the Parkes telescope at 3.3\,GHz, yielding a map that exhibits strong morphological resemblance to maps presented in Carretti et al. (2013; Figure 1). This is presented in Figure \ref{fig:a3667_1372sub_uvtaper}.

There is an extension of emission toward the cluster centre seen in the \emph{uv}-tapered map (see Figure \ref{fig:a3667_1372sub_uvtaper}) which could indicate the presence of a radio bridge. However, examination of the wide-field ATCA image at 1400 MHz reveals 3 point sources in the region of this extension, all exceeding $6\sigma$ significance. These were either not detected in the 2.3--3.0\,GHz frequency range imaged in \cite{2013MNRAS.430.1414C} or below the 3 mJy threshold of the catalog. It is possible that these sources contribute the flux recovered by KAT-7, as the emission in this region is only detected at the $3\sigma$ level.

\section{Discussion}\label{sec:DISC}
The residual emission seen in Figure \ref{fig:a3667_1372sub} lies spatially co-located with B2007-569. We do not have enough beams across the source to fully determine its size, but it appears extended in nature with an apparent major axis of approximately 500 arcsec; the deconvolved size is 270 arcsec. This would correspond to a physical size of 278 kpc, consistent with the scale size of known mini-haloes (typically $150-500$\,kpc in size, e.g. \citealt{2013A&A...557A..99K}). One explanation is that the residual emission might be attributed to flux from the tails of B2007-569 that exists on scales unrecovered by the ATCA observations. However, the archival ATCA data were taken in the 1.5D configuration, with baselines in the range 105\,m to 4.4\,km. At 1400\,MHz therefore, the maximum scale size recovered by these data will be of the order of 7 arcmin, which overlaps well with the scale size range recovered by KAT-7. Hence the ATCA observations should be missing little flux from B2007-569 unless there exists some power on very large angular scales. An alternative explanation is that this residual emission is a new candidate radio mini-halo.

\cite{2014MNRAS.445..330H} have shown that A3667 lies coincident with a region of very extended Galactic emission. From this work we cannot rule out the possibility that the recovered emission we tentatively identify as a mini-halo is the result of poorly-sampled Galactic emission. However, from the Galactic synchrotron analysis of \cite{2008A&A...479..641L} and following \cite{2013MNRAS.430.1414C} we estimate the contamination at the 1372 MHz beam scale of KAT-7 (4.19 arcmin) to be $\sim4.5$ mJy. This contribution is small compared to the flux density recovered from our MHc (which peaks at $70.8\pm4.5$ mJy beam$^{-1}$).

The suggestion of a radio halo in A3667 is not a new idea. Johnston-Hollitt (2003 and 2004) report a candidate radio halo with an integrated flux density of $33\pm6$ mJy at 1.4GHz. Additionally, \cite{2013MNRAS.430.1414C} suggest a radio halo of integrated flux $44\pm6$ mJy at 3.3 GHz. However, both these candidate haloes are coincident with the peak of X-ray emission, whereas this mini-halo candidate (MHc) lies approximately coincident with B2007-569, the brightest cluster galaxy (BCG). The co-location of the MHc and BCG is consistent with previous mini-halo (MH) detections.

\cite{2014ApJ...781....9G} present the most recent catalog of known MH, listing a total of 15 confirmed MH, with 3 additional MHc and a further 3 whose classification as MH is uncertain. Taking radio power from Giacintucci et al. (2014; Table 5, confirmed/candidates only) and bolometric X-ray luminosity from the Archive of Chandra Cluster Entropy Profile Tables (ACCEPT; \citealt{2008ApJ...683L.107C}) we investigate whether the proposed MHc in A3667 is consistent with trends in the larger MH population.

\begin{figure*}
	\begin{subfigure}[l]{0.497\textwidth}
		\includegraphics[width=\textwidth]{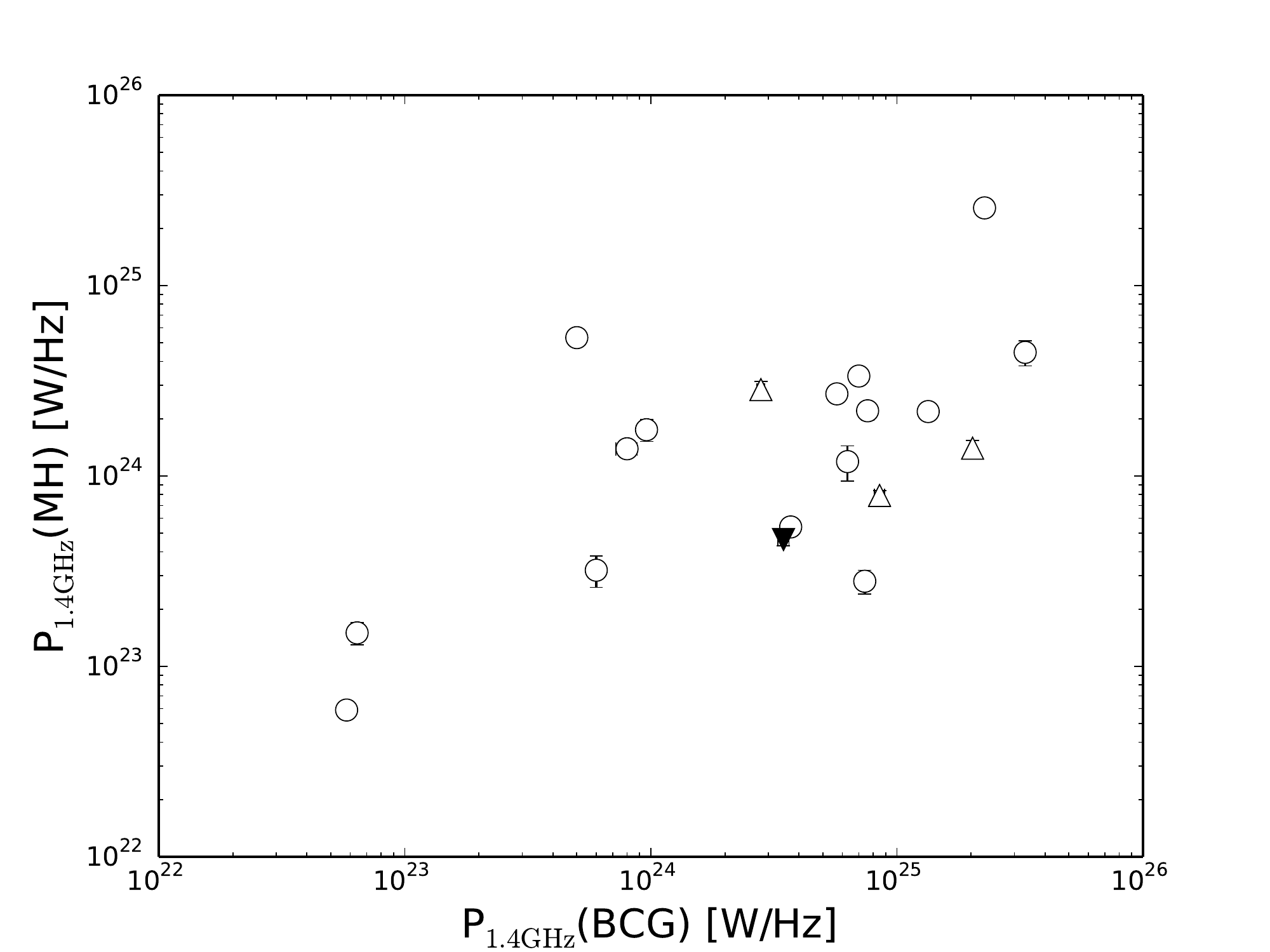}
		\label{fig:bcg_mh}
	\end{subfigure}
	\begin{subfigure}[r]{0.497\textwidth}
		{\includegraphics[width=\textwidth]{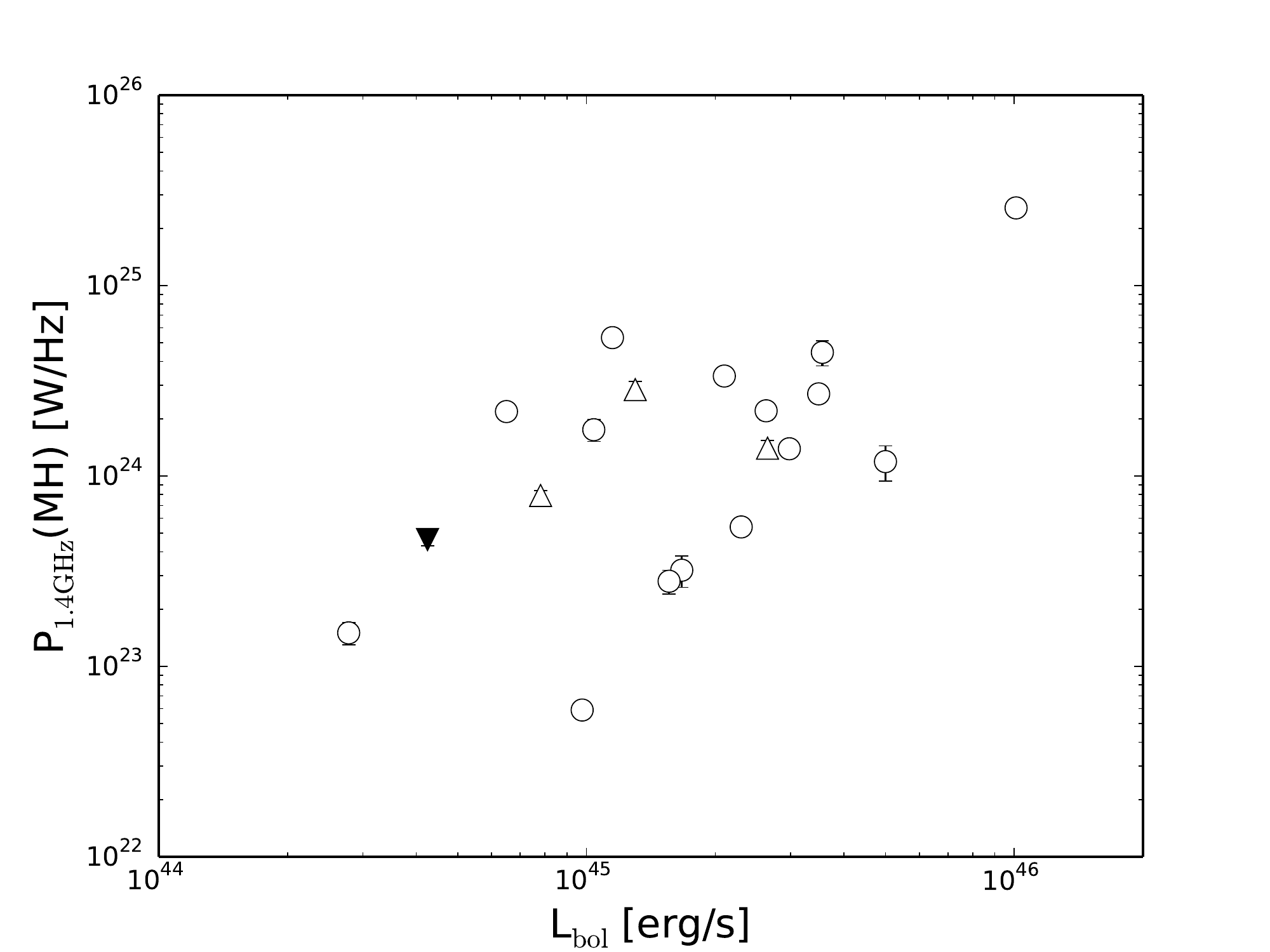}}
		\label{fig:lbol_mh}
	\end{subfigure}
\caption{Scaling trends for clusters with confirmed or candidate MH. \textsc{left panel} shows the trend in MH and BCG radio power scaling; \textsc{right panel} shows the scaling trend between cluster X-ray bolometric luminosity and MH radio power. Unfilled circles are clusters with confirmed MH, unfilled triangles are clusters with candidate MH \protect\citep{2014ApJ...781....9G}; filled symbols correspond to values for A3667. All cluster L$_{\rm{bol}}$ values are from ACCEPT \protect\citep{2008ApJ...683L.107C}; cluster MH/BCG radio power are quoted from \protect\cite{2014ApJ...781....9G} with the exception of A3667, where the MHc/BCG radio power were determined in this paper and have here been scaled to match the cosmology of \protect\cite{2014ApJ...781....9G}. Errors in measured radio power are included in the plot.} 
\label{fig:mh_scaling_relations}
\end{figure*}

This new MHc possesses an integrated flux density of 67.2$\pm4.9$ mJy at 1372 MHz, which corresponds to a radio power of P$_{\rm{1.4\,\,\,GHz}}=4.28\pm0.31\times10^{23}$ W Hz$^{-1}$ with our cosmology. Our 1400\,MHz ATCA map of B2007-569 has an integrated flux density of $497\pm25$ mJy, corresponding to a radio power of P$_{\rm{1.4\,\,\,GHz}}=3.39\pm0.17\times10^{24}$ W Hz$^{-1}$. Despite large scatter, from Figure \ref{fig:mh_scaling_relations}, A3667 appears consistent with the trend that clusters which host more powerful MH tend to host more powerful BCGs (e.g. \citealt{2009A&A...499..371G}; \citealt{2014ApJ...781....9G}). Additionally, A3667 is approximately consistent with the trend that the more powerful MH tend to be hosted by more X-ray bright clusters. Note that in Figure \ref{fig:mh_scaling_relations} we have scaled the radio power of the MHc in A3667 to match the cosmology of \cite{2014ApJ...781....9G}.

\cite{2014MNRAS.445..330H} subtract B2007-569 from their 226 MHz data by modelling it as a point source. After subtraction, Hindson et al. recover emission at the level $\sim0.50\pm0.05$ Jy beam$^{-1}$ $\pm10.8\%$. In estimating this error we quote the image rms (50 mJy beam$^{-1}$) and the approximate calibration uncertainty ($\sim10.8$ per cent) from \cite{2014MNRAS.445..330H}. The flux density recovered from this new MHc peaks at $70.8\pm4.5$ mJy beam$^{-1}$, equivalent to a surface brightness of approximately 1.1 $\mu$Jy arcsec$^{-2}$. A spectral index of $\alpha_{\rm{MH}}=-0.85$ would yield a 226 MHz flux density consistent with that recovered by Hindson et al., although this is slightly flatter than expected from observations. 

A small sample of cool-core clusters host radio relics, but these are generally believed to arise as a result of weak shocks or off-axis mergers \citep{2012A&ARv..20...54F}. On the other hand, while it is a confirmed cool-core cluster (albeit a weak cool-core; \citealt{2010A&A...513A..37H}) the merger event in A3667 is a major merger, with the sub-clusters colliding approximately head-on. It has been suggested however, that A3667 is observed in a post-merger state \citep{2010ApJ...715.1143F}. Two questions arise: (1) could a cool-core in A3667 survive a major merger event and (2) if a cool-core cannot survive a major merger, has sufficient time elapsed for re-establishment of a cool-core?

Based on a selection of clusters from the Giant Metrewave Radio Telescope (GMRT) Radio Halo Survey (GRHS; see \citealt{2007A&A...463..937V}; \citealt{2008A&A...484..327V} for details) \cite{2011A&A...532A.123R} derive the time-scale ($t_{\rm{NCC}\rightarrow\rm{CC}}$) on which non-cool core clusters relax and re-establish cool cores. They derive an upper limit of 1.7 Gyr, and time-scales in the range 1.2$-$2.7 Gyr with the inclusion of clusters in the HIFLUGCS subsample of the ACCEPT catalog that were not observed during the GRHS. This is consistent with expectations based on the estimations of radiative lifetime of radio haloes, $t_{\rm{RH}}\sim1$ Gyr \citep{2009A&A...507..661B}. This relaxation time-scale is significantly shorter than the expected cooling lifetime of non-cool-core clusters based on thermodynamic arguments, which is of the order of 10 Gyr \citep{2010A&A...510A..83R}. Based on X-ray observations, the structure in the SE cold front in A3667 bears close resemblance to the simulated (\citealt{2003MNRAS.346...13H}) structure at 3--4 Gyr post core-passage \citep{2004A&A...426....1B}. While this would be consistent with such a timescale, a newly-formed cool core would likely lead to a detectable cooling flow; however observations have not suggested the presence of a cooling flow in A3667.

Additionally, in clusters which possess cooling flows the cool-core is typically found coincident with the peak of the X-ray emission, whereas the MHc and BCG in A3667 are co-located with the secondary sub-cluster, and significantly offset from the X-ray emission peak. Previous X-ray observations indicate the presence of a cool region of gas co-located with the BCG (Markevitch et al. 1999) which could indicate a relic cool-core, though further observations are required to confirm this. Some simulations have suggested that major merger events occurring in the local Universe have difficulty destroying established cool cores \citep{2008ApJ...675.1125B}; however the question of the extent to which merger events disrupt cool cores is still open.

Given the co-location of the BCG (and MHc) in A3667 with the secondary sub-cluster (travelling in a NW direction; \citealt{2009ApJ...693..901O}) offset from the peak of the X-ray emission, and lack of observed cooling flow, we suggest that this would likely not be a newly-formed cool-core, but a relic cool-core which has survived the cluster merger. In this scenario, the likely source of electron acceleration would be gas sloshing caused by perturbation of the cool-core by the merger event (e.g. ZuHone et al. 2011).

\section{Conclusions}\label{sec:CONC}

In this work, we have mapped the dual radio relics of the galaxy cluster Abell 3667 in both total intensity and linearly-polarized intensity at 1826 and 1372 MHz with KAT-7. We recover integrated flux densities and polarization fractions that are broadly consistent with previous work at similar frequencies.

Combining our new KAT-7 data with ancillary data from SUMSS we have mapped the spectral profile of both radio relics between 1826 and 843 MHz. We find typical spectral index values of $\alpha = -0.8\pm0.2 \, \,  (-0.5\pm0.2)$ in the North-West (South-East) relic. The North-West relic exhibits a rough transverse gradient, steepening toward the cluster centre and flattening toward the outer edge, indicative of differences in the age of the electrons responsible for the emission.

The typical spectral index of the South-East relic is consistent with the predicted spectral index according to the diffusive shock acceleration mechanism. Our polarization analysis is additionally consistent with this view, as the magnetic field vectors align along the long axis of the South-East relic. The polarization analysis also suggests that the merger axis may be slightly inclined, as the magnetic field vectors for the North-West relic do not quite lie along the long relic axis, indicative of Faraday rotation along the line-of-sight to this relic that does not occur for the South-East relic.

Using the spectral index map in conjunction with our KAT-7 map at 1826 MHz, we have estimated the equipartition magnetic field strength ($B_{\rm{eq}}$) for the NW relic using the formalism of \cite{2005AN....326..414B}. We find typical values of $B_{\rm{eq}} = 1.5-3 \,\,\mu$G, consistent with previous estimates from equipartition arguments.

We have subtracted point sources identified in previous work from the North-West relic region to isolate flux density contribution from the diffuse emission. Additionally, we have subtracted the head-tail radio galaxy MRC B2007-569 by using archival ATCA observations to produce a model using multi-frequency synthesis across the wide bandwidth available with the CABB. Localised residual emission remains, spatially co-located with MRC B2007-569, which we have tentatively identified as a MHc. This MHc has an integrated flux density of $67.2\pm4.9$ mJy at 1.37 GHz, corresponding to P$_{\rm{1.4\,\,\,GHz}}=4.28\pm0.31\times10^{23}$ W Hz$^{-1}$. Subsequent investigation has shown this to be consistent with previously-established trends.

We suggest that the turbulence required for electron reacceleration may be generated by gas sloshing set off by the major merger perturbing (but not destroying) the weak cool-core in A3667 (see \citealt{2010A&A...513A..37H} for the classification of a weak cool-core cluster). This scenario is consistent with some simulations which have shown that established cool-cores in the local Universe are difficult to destroy with major merger events \citep{2008ApJ...675.1125B}.


\subsection{Acknowledgements}
This work was carried out as part of commissioning efforts of KAT-7. All KAT-7 data and images are provided courtesy of SKA SA - http://www.ska.ac.za. We would like to thank the Square Kilometre Array South Africa (SKA SA) collaboration, and recognise the efforts of the KAT-7 observers and engineers for making this work possible. We extend our thanks to our referee M.~Johnston-Hollitt for detailed comments and suggestions that assisted in the interpretation of our results and improved our scientific output. CJR gratefully acknowledges both support from the UK Science \& Technology Facilities Council (STFC) and the support from SKA SA that enabled this work to be conducted. AMS gratefully acknowledges support from the European Research Council under grant  ERC-2012-StG-307215 LODESTONE.

CJR wishes to thank T.~Mauch and L.~Richter for helpful discussions during the data reduction process; and M.~Johnston-Hollitt for informative discussions during results analysis. This work made use of the cosmology calculator developed by Wright (2006) and the Archive of Chandra Cluster Entropy Profile Tables (ACCEPT; \citealt{2008ApJ...683L.107C}). The colour schemes for Figures \ref{fig:spix}, \ref{fig:beq} and \ref{fig:b2007_modelling} were made using the cubehelix colour scheme generator developed by D.~Green  \citep{2011BASI...39..289G} to generate a custom colour scheme specifically for this work.

\bibliographystyle{mn2e}
\bibliography{ACO3667_riseley_v3}

\begin{thebibliography}{74}
\expandafter\ifx\csname natexlab\endcsname\relax\def\natexlab#1{#1}\fi

\bibitem[{{Akamatsu} {et~al}\mbox{.}(2012{\natexlab{a}}){Akamatsu}, {de Plaa},
  {Kaastra}, {Ishisaki}, {Ohashi}, {Kawaharada}, \&
  {Nakazawa}}]{2012PASJ...64...49A}
{Akamatsu} H., {de Plaa} J., {Kaastra} J., {Ishisaki} Y., {Ohashi} T.,
  {Kawaharada} M., {Nakazawa} K., 2012{\natexlab{a}}, \pasj, 64, 49

\bibitem[{{Akamatsu} {et~al}\mbox{.}(2012{\natexlab{b}}){Akamatsu}, {Takizawa},
  {Nakazawa}, {Fukazawa}, {Ishisaki}, \& {Ohashi}}]{2012PASJ...64...67A}
{Akamatsu} H., {Takizawa} M., {Nakazawa} K., {Fukazawa} Y., {Ishisaki} Y.,
  {Ohashi} T., 2012{\natexlab{b}}, \pasj, 64, 67

\bibitem[{{Armstrong} {et~al}\mbox{.}(2013){Armstrong}, {Fender}, {Nicolson},
  {Ratcliffe}, {Linares}, {Horrell}, {Richter}, {Schurch}, {Coriat}, {Woudt},
  {Jonas}, {Booth}, \& {Fanaroff}}]{2013MNRAS.433.1951A}
{Armstrong} R.~P. {et~al.}, 2013, \mnras, 433, 1951

\bibitem[{{Baars} {et~al}\mbox{.}(1977){Baars}, {Genzel}, {Pauliny-Toth}, \&
  {Witzel}}]{1977A&A....61...99B}
{Baars} J.~W.~M., {Genzel} R., {Pauliny-Toth} I.~I.~K., {Witzel} A., 1977,
  \aap, 61, 99

\bibitem[{{Beck} \& {Krause}(2005)}]{2005AN....326..414B}
{Beck} R., {Krause} M., 2005, Astronomische Nachrichten, 326, 414

\bibitem[{{Bock}, {Large} \& {Sadler}(1999){Bock}, {Large}, \&
  {Sadler}}]{1999AJ....117.1578B}
{Bock} D.~C.-J., {Large} M.~I., {Sadler} E.~M., 1999, \aj, 117, 1578

\bibitem[{{Brentjens} \& {de Bruyn}(2005)}]{2005A&A...441.1217B}
{Brentjens} M.~A., {de Bruyn} A.~G., 2005, \aap, 441, 1217

\bibitem[{{Briel}, {Finoguenov} \& {Henry}(2004){Briel}, {Finoguenov}, \&
  {Henry}}]{2004A&A...426....1B}
{Briel} U.~G., {Finoguenov} A., {Henry} J.~P., 2004, \aap, 426, 1

\bibitem[{{Brunetti} {et~al}\mbox{.}(2009){Brunetti}, {Cassano}, {Dolag}, \&
  {Setti}}]{2009A&A...507..661B}
{Brunetti} G., {Cassano} R., {Dolag} K., {Setti} G., 2009, \aap, 507, 661

\bibitem[{{Burns} {et~al}\mbox{.}(2008){Burns}, {Hallman}, {Gantner}, {Motl},
  \& {Norman}}]{2008ApJ...675.1125B}
{Burns} J.~O., {Hallman} E.~J., {Gantner} B., {Motl} P.~M., {Norman} M.~L.,
  2008, \apj, 675, 1125

\bibitem[{{Carretti} {et~al}\mbox{.}(2013){Carretti}, {Brown},
  {Staveley-Smith}, {Malarecki}, {Bernardi}, {Gaensler}, {Haverkorn},
  {Kesteven}, \& {Poppi}}]{2013MNRAS.430.1414C}
{Carretti} E. {et~al.}, 2013, \mnras, 430, 1414

\bibitem[{{Cassano}, {Gitti} \& {Brunetti}(2008){Cassano}, {Gitti}, \&
  {Brunetti}}]{2008A&A...486L..31C}
{Cassano} R., {Gitti} M., {Brunetti} G., 2008, \aap, 486, L31

\bibitem[{{Cavagnolo} {et~al}\mbox{.}(2008){Cavagnolo}, {Donahue}, {Voit}, \&
  {Sun}}]{2008ApJ...683L.107C}
{Cavagnolo} K.~W., {Donahue} M., {Voit} G.~M., {Sun} M., 2008, \apjl, 683, L107

\bibitem[{{de Gasperin} {et~al}\mbox{.}(2014){de Gasperin}, {van Weeren},
  {Bruggen}, {Vazza}, {Bonafede}, \& {Intema}}]{2014arXiv1408.2677D}
{de Gasperin} F., {van Weeren} R.~J., {Bruggen} M., {Vazza} F., {Bonafede} A.,
  {Intema} H.~T., 2014, ArXiv e-prints

\bibitem[{{de Zotti} {et~al}\mbox{.}(2010){de Zotti}, {Massardi}, {Negrello},
  \& {Wall}}]{2010A&ARv..18....1D}
{de Zotti} G., {Massardi} M., {Negrello} M., {Wall} J., 2010, \aapr, 18, 1

\bibitem[{{Dennison}(1980)}]{1980ApJ...239L..93D}
{Dennison} B., 1980, \apjl, 239, L93

\bibitem[{{Feretti}(2005)}]{2005AdSpR..36..729F}
{Feretti} L., 2005, Advances in Space Research, 36, 729

\bibitem[{{Feretti} {et~al}\mbox{.}(2012){Feretti}, {Giovannini}, {Govoni}, \&
  {Murgia}}]{2012A&ARv..20...54F}
{Feretti} L., {Giovannini} G., {Govoni} F., {Murgia} M., 2012, \aapr, 20, 54

\bibitem[{{Ferrari} {et~al}\mbox{.}(2008){Ferrari}, {Govoni}, {Schindler},
  {Bykov}, \& {Rephaeli}}]{2008SSRv..134...93F}
{Ferrari} C., {Govoni} F., {Schindler} S., {Bykov} A.~M., {Rephaeli} Y., 2008,
  \ssr, 134, 93

\bibitem[{{Finoguenov} {et~al}\mbox{.}(2010){Finoguenov}, {Sarazin},
  {Nakazawa}, {Wik}, \& {Clarke}}]{2010ApJ...715.1143F}
{Finoguenov} A., {Sarazin} C.~L., {Nakazawa} K., {Wik} D.~R., {Clarke} T.~E.,
  2010, \apj, 715, 1143

\bibitem[{{Giacintucci} {et~al}\mbox{.}(2014){Giacintucci}, {Markevitch},
  {Venturi}, {Clarke}, {Cassano}, \& {Mazzotta}}]{2014ApJ...781....9G}
{Giacintucci} S., {Markevitch} M., {Venturi} T., {Clarke} T.~E., {Cassano} R.,
  {Mazzotta} P., 2014, \apj, 781, 9

\bibitem[{{Gitti} {et~al}\mbox{.}(2004){Gitti}, {Brunetti}, {Feretti}, \&
  {Setti}}]{2004A&A...417....1G}
{Gitti} M., {Brunetti} G., {Feretti} L., {Setti} G., 2004, \aap, 417, 1

\bibitem[{{Gitti}, {Brunetti} \& {Setti}(2002){Gitti}, {Brunetti}, \&
  {Setti}}]{2002A&A...386..456G}
{Gitti} M., {Brunetti} G., {Setti} G., 2002, \aap, 386, 456

\bibitem[{{Goss} {et~al}\mbox{.}(1982){Goss}, {Ekers}, {Skellern}, \&
  {Smith}}]{1982MNRAS.198..259G}
{Goss} W.~M., {Ekers} R.~D., {Skellern} D.~J., {Smith} R.~M., 1982, \mnras,
  198, 259

\bibitem[{{Govoni} {et~al}\mbox{.}(2001){Govoni}, {En{\ss}lin}, {Feretti}, \&
  {Giovannini}}]{2001A&A...369..441G}
{Govoni} F., {En{\ss}lin} T.~A., {Feretti} L., {Giovannini} G., 2001, \aap,
  369, 441

\bibitem[{{Govoni} {et~al}\mbox{.}(2009){Govoni}, {Murgia}, {Markevitch},
  {Feretti}, {Giovannini}, {Taylor}, \& {Carretti}}]{2009A&A...499..371G}
{Govoni} F., {Murgia} M., {Markevitch} M., {Feretti} L., {Giovannini} G.,
  {Taylor} G.~B., {Carretti} E., 2009, \aap, 499, 371

\bibitem[{{Green}(2007)}]{2007BASI...35...77G}
{Green} D.~A., 2007, Bulletin of the Astronomical Society of India, 35, 77

\bibitem[{{Green}(2011)}]{2011BASI...39..289G}
{Green} D.~A., 2011, Bulletin of the Astronomical Society of India, 39, 289

\bibitem[{{Gutierrez} \& {Krawczynski}(2005)}]{2005ApJ...619..161G}
{Gutierrez} K., {Krawczynski} H., 2005, \apj, 619, 161

\bibitem[{{Heald}(2009)}]{2009IAUS..259..591H}
{Heald} G., 2009, in IAU Symposium, Vol. 259, IAU Symposium, {Strassmeier}
  K.~G., {Kosovichev} A.~G., {Beckman} J.~E., eds., pp. 591--602

\bibitem[{{Heinz} {et~al}\mbox{.}(2003){Heinz}, {Churazov}, {Forman}, {Jones},
  \& {Briel}}]{2003MNRAS.346...13H}
{Heinz} S., {Churazov} E., {Forman} W., {Jones} C., {Briel} U.~G., 2003,
  \mnras, 346, 13

\bibitem[{{Hindson} {et~al}\mbox{.}(2014){Hindson}, {Johnston-Hollitt},
  {Hurley-Walker}, {Buckley}, {Morgan}, {Carretti}, {Dwarakanath}, {Bell},
  {Bernardi}, {Bhat}, {Bowman}, {Briggs}, {Cappallo}, {Corey}, {Deshpande},
  {Emrich}, {Ewall-Wice}, {Feng}, {Gaensler}, {Goeke}, {Greenhill}, {Hazelton},
  {Jacobs}, {Kaplan}, {Kasper}, {Kratzenberg}, {Kudryavtseva}, {Lenc},
  {Lonsdale}, {Lynch}, {McWhirter}, {McKinley}, {Mitchell}, {Morales},
  {Morgan}, {Oberoi}, {Ord}, {Pindor}, {Prabu}, {Procopio}, {Offringa},
  {Riding}, {Rogers}, {Roshi}, {Shankar}, {Srivani}, {Subrahmanyan}, {Tingay},
  {Waterson}, {Wayth}, {Webster}, {Whitney}, {Williams}, \&
  {Williams}}]{2014MNRAS.445..330H}
{Hindson} L. {et~al.}, 2014, \mnras, 445, 330

\bibitem[{{Hudson} {et~al}\mbox{.}(2010){Hudson}, {Mittal}, {Reiprich},
  {Nulsen}, {Andernach}, \& {Sarazin}}]{2010A&A...513A..37H}
{Hudson} D.~S., {Mittal} R., {Reiprich} T.~H., {Nulsen} P.~E.~J., {Andernach}
  H., {Sarazin} C.~L., 2010, \aap, 513, A37

\bibitem[{{Hunstead}(1991)}]{1991AuJPh..44..743H}
{Hunstead} R.~W., 1991, Australian Journal of Physics, 44, 743

\bibitem[{{Hurley-Walker} {et~al}\mbox{.}(2014){Hurley-Walker}, {Morgan},
  {Wayth}, {Hancock}, {Bell}, {Bernardi}, {Bhat}, {Briggs}, {Deshpande},
  {Ewall-Wice}, {Feng}, {Hazelton}, {Hindson}, {Jacobs}, {Kaplan Nadia
  Kudryavtseva}, {Lenc}, {McKinley}, {Mitchell}, {Pindor}, {Procopio},
  {Oberoi}, {Offringa}, {Ord}, {Riding}, {Bowman}, {Cappallo}, {Corey},
  {Emrich}, {Gaensler}, {Goeke}, {Greenhill}, {Hewitt}, {Johnston-Hollitt},
  {Kasper}, {Kratzenberg}, {Lonsdale}, {Lynch}, {McWhirter}, {Morales},
  {Morgan}, {Prabu}, {Rogers}, {Roshi}, {Shankar}, {Srivani}, {Subrahmanyan},
  {Tingay}, {Waterson}, {Webster}, {Whitney}, {Williams}, \&
  {Williams}}]{2014arXiv1410.0790H}
{Hurley-Walker} N. {et~al.}, 2014, ArXiv e-prints

\bibitem[{{Johnston-Hollitt}(2003)}]{2003PhDT.........3J}
{Johnston-Hollitt} M., 2003, PhD thesis, University of Adelaide

\bibitem[{{Johnston-Hollitt}(2004)}]{2004rcfg.proc...51J}
{Johnston-Hollitt} M., 2004, in The Riddle of Cooling Flows in Galaxies and
  Clusters of galaxies, {Reiprich} T., {Kempner} J., {Soker} N., eds., p.~51

\bibitem[{{Johnston-Hollitt}, {Hunstead} \& {Corbett}(2008){Johnston-Hollitt},
  {Hunstead}, \& {Corbett}}]{2008A&A...479....1J}
{Johnston-Hollitt} M., {Hunstead} R.~W., {Corbett} E., 2008, \aap, 479, 1

\bibitem[{{Kale} {et~al}\mbox{.}(2013){Kale}, {Venturi}, {Giacintucci},
  {Dallacasa}, {Cassano}, {Brunetti}, {Macario}, \&
  {Athreya}}]{2013A&A...557A..99K}
{Kale} R., {Venturi} T., {Giacintucci} S., {Dallacasa} D., {Cassano} R.,
  {Brunetti} G., {Macario} G., {Athreya} R., 2013, \aap, 557, A99

\bibitem[{{Kempner} {et~al}\mbox{.}(2004){Kempner}, {Blanton}, {Clarke},
  {En{\ss}lin}, {Johnston-Hollitt}, \& {Rudnick}}]{2004rcfg.proc..335K}
{Kempner} J.~C., {Blanton} E.~L., {Clarke} T.~E., {En{\ss}lin} T.~A.,
  {Johnston-Hollitt} M., {Rudnick} L., 2004, in The Riddle of Cooling Flows in
  Galaxies and Clusters of galaxies, {Reiprich} T., {Kempner} J., {Soker} N.,
  eds., p. 335

\bibitem[{{La Porta} {et~al}\mbox{.}(2008){La Porta}, {Burigana}, {Reich}, \&
  {Reich}}]{2008A&A...479..641L}
{La Porta} L., {Burigana} C., {Reich} W., {Reich} P., 2008, \aap, 479, 641

\bibitem[{{Large} {et~al}\mbox{.}(1981){Large}, {Mills}, {Little}, {Crawford},
  \& {Sutton}}]{1981MNRAS.194..693L}
{Large} M.~I., {Mills} B.~Y., {Little} A.~G., {Crawford} D.~F., {Sutton} J.~M.,
  1981, \mnras, 194, 693

\bibitem[{{Macario} {et~al}\mbox{.}(2011){Macario}, {Markevitch},
  {Giacintucci}, {Brunetti}, {Venturi}, \& {Murray}}]{2011ApJ...728...82M}
{Macario} G., {Markevitch} M., {Giacintucci} S., {Brunetti} G., {Venturi} T.,
  {Murray} S.~S., 2011, \apj, 728, 82

\bibitem[{{Markevitch}, {Sarazin} \& {Vikhlinin}(1999){Markevitch}, {Sarazin},
  \& {Vikhlinin}}]{1999ApJ...521..526M}
{Markevitch} M., {Sarazin} C.~L., {Vikhlinin} A., 1999, \apj, 521, 526

\bibitem[{{Mauch} {et~al}\mbox{.}(2003){Mauch}, {Murphy}, {Buttery}, {Curran},
  {Hunstead}, {Piestrzynski}, {Robertson}, \& {Sadler}}]{2003MNRAS.342.1117M}
{Mauch} T., {Murphy} T., {Buttery} H.~J., {Curran} J., {Hunstead} R.~W.,
  {Piestrzynski} B., {Robertson} J.~G., {Sadler} E.~M., 2003, \mnras, 342, 1117

\bibitem[{{Nakazawa} {et~al}\mbox{.}(2009){Nakazawa}, {Sarazin}, {Kawaharada},
  {Kitaguchi}, {Okuyama}, {Makishima}, {Kawano}, {Fukazawa}, {Inoue},
  {Takizawa}, {Wik}, {Finoguenov}, \& {Clarke}}]{2009PASJ...61..339N}
{Nakazawa} K. {et~al.}, 2009, \pasj, 61, 339

\bibitem[{{Oppermann} {et~al}\mbox{.}(2012){Oppermann}, {Junklewitz},
  {Robbers}, {Bell}, {En{\ss}lin}, {Bonafede}, {Braun}, {Brown}, {Clarke},
  {Feain}, {Gaensler}, {Hammond}, {Harvey-Smith}, {Heald}, {Johnston-Hollitt},
  {Klein}, {Kronberg}, {Mao}, {McClure-Griffiths}, {O'Sullivan}, {Pratley},
  {Robishaw}, {Roy}, {Schnitzeler}, {Sotomayor-Beltran}, {Stevens}, {Stil},
  {Sunstrum}, {Tanna}, {Taylor}, \& {Van Eck}}]{2012A&A...542A..93O}
{Oppermann} N. {et~al.}, 2012, \aap, 542, A93

\bibitem[{{Owers}, {Couch} \& {Nulsen}(2009){Owers}, {Couch}, \&
  {Nulsen}}]{2009ApJ...693..901O}
{Owers} M.~S., {Couch} W.~J., {Nulsen} P.~E.~J., 2009, \apj, 693, 901

\bibitem[{{Owers} {et~al}\mbox{.}(2009){Owers}, {Nulsen}, {Couch}, \&
  {Markevitch}}]{2009ApJ...704.1349O}
{Owers} M.~S., {Nulsen} P.~E.~J., {Couch} W.~J., {Markevitch} M., 2009, \apj,
  704, 1349

\bibitem[{{Perley} \& {Butler}(2013)}]{2013ApJS..206...16P}
{Perley} R.~A., {Butler} B.~J., 2013, \apjs, 206, 16

\bibitem[{{Pfrommer} \& {En{\ss}lin}(2004)}]{2004MNRAS.352...76P}
{Pfrommer} C., {En{\ss}lin} T.~A., 2004, \mnras, 352, 76

\bibitem[{{Rau} \& {Cornwell}(2011)}]{2011A&A...532A..71R}
{Rau} U., {Cornwell} T.~J., 2011, \aap, 532, A71

\bibitem[{{Rayner}, {Norris} \& {Sault}(2000){Rayner}, {Norris}, \&
  {Sault}}]{2000MNRAS.319..484R}
{Rayner} D.~P., {Norris} R.~P., {Sault} R.~J., 2000, \mnras, 319, 484

\bibitem[{{Reiprich} \& {B{\"o}hringer}(2002)}]{2002ApJ...567..716R}
{Reiprich} T.~H., {B{\"o}hringer} H., 2002, \apj, 567, 716

\bibitem[{{Reynolds}(1994)}]{reynolds94}
{Reynolds} J.~E., 1994, {A Revised Flux Scale for the AT Compact Array, Tech.
  Rep. 39.3/040, Australia Telescope National Facility -- }. 1997, ATCA
  Calibrator Source Catalogue, ftp://ftp.atnf.csiro.au/pub/atnfdo
  cs/guides/at.cat

\bibitem[{{Rossetti} {et~al}\mbox{.}(2011){Rossetti}, {Eckert}, {Cavalleri},
  {Molendi}, {Gastaldello}, \& {Ghizzardi}}]{2011A&A...532A.123R}
{Rossetti} M., {Eckert} D., {Cavalleri} B.~M., {Molendi} S., {Gastaldello} F.,
  {Ghizzardi} S., 2011, \aap, 532, A123

\bibitem[{{Rossetti} \& {Molendi}(2010)}]{2010A&A...510A..83R}
{Rossetti} M., {Molendi} S., 2010, \aap, 510, A83

\bibitem[{{R\"{o}ttgering} {et~al}\mbox{.}(1997){R\"{o}ttgering}, {Wieringa},
  {Hunstead}, \& {Ekers}}]{1997MNRAS.290..577R}
{R\"{o}ttgering} H.~J.~A., {Wieringa} M.~H., {Hunstead} R.~W., {Ekers} R.~D.,
  1997, \mnras, 290, 577

\bibitem[{{Sault}, {Teuben} \& {Wright}(1995){Sault}, {Teuben}, \&
  {Wright}}]{1995ASPC...77..433S}
{Sault} R.~J., {Teuben} P.~J., {Wright} M.~C.~H., 1995, in Astronomical Society
  of the Pacific Conference Series, Vol.~77, Astronomical Data Analysis
  Software and Systems IV, {Shaw} R.~A., {Payne} H.~E., {Hayes} J.~J.~E., eds.,
  p. 433

\bibitem[{{Schenck} {et~al}\mbox{.}(2014){Schenck}, {Datta}, {Burns}, \&
  {Skillman}}]{2014AJ....148...23S}
{Schenck} D.~E., {Datta} A., {Burns} J.~O., {Skillman} S., 2014, \aj, 148, 23

\bibitem[{{Slee} {et~al}\mbox{.}(2001){Slee}, {Roy}, {Murgia}, {Andernach}, \&
  {Ehle}}]{2001AJ....122.1172S}
{Slee} O.~B., {Roy} A.~L., {Murgia} M., {Andernach} H., {Ehle} M., 2001, \aj,
  122, 1172

\bibitem[{{Sodre} {et~al}\mbox{.}(1992){Sodre}, {Capelato}, {Steiner},
  {Proust}, \& {Mazure}}]{1992MNRAS.259..233S}
{Sodre}, Jr. L., {Capelato} H.~V., {Steiner} J.~E., {Proust} D., {Mazure} A.,
  1992, \mnras, 259, 233

\bibitem[{{Tingay} {et~al}\mbox{.}(2013){Tingay}, {Goeke}, {Bowman}, {Emrich},
  {Ord}, {Mitchell}, {Morales}, {Booler}, {Crosse}, {Wayth}, {Lonsdale},
  {Tremblay}, {Pallot}, {Colegate}, {Wicenec}, {Kudryavtseva}, {Arcus},
  {Barnes}, {Bernardi}, {Briggs}, {Burns}, {Bunton}, {Cappallo}, {Corey},
  {Deshpande}, {Desouza}, {Gaensler}, {Greenhill}, {Hall}, {Hazelton}, {Herne},
  {Hewitt}, {Johnston-Hollitt}, {Kaplan}, {Kasper}, {Kincaid}, {Koenig},
  {Kratzenberg}, {Lynch}, {Mckinley}, {Mcwhirter}, {Morgan}, {Oberoi},
  {Pathikulangara}, {Prabu}, {Remillard}, {Rogers}, {Roshi}, {Salah}, {Sault},
  {Udaya-Shankar}, {Schlagenhaufer}, {Srivani}, {Stevens}, {Subrahmanyan},
  {Waterson}, {Webster}, {Whitney}, {Williams}, {Williams}, \&
  {Wyithe}}]{2013PASA...30....7T}
{Tingay} S.~J. {et~al.}, 2013, PASA, 30, 7

\bibitem[{{van Weeren} {et~al}\mbox{.}(2010){van Weeren}, {R{\"o}ttgering},
  {Br{\"u}ggen}, \& {Hoeft}}]{2010Sci...330..347V}
{van Weeren} R.~J., {R{\"o}ttgering} H.~J.~A., {Br{\"u}ggen} M., {Hoeft} M.,
  2010, Science, 330, 347

\bibitem[{{van Weeren} {et~al}\mbox{.}(2012){van Weeren}, {R{\"o}ttgering},
  {Intema}, {Rudnick}, {Br{\"u}ggen}, {Hoeft}, \& {Oonk}}]{2012A&A...546A.124V}
{van Weeren} R.~J., {R{\"o}ttgering} H.~J.~A., {Intema} H.~T., {Rudnick} L.,
  {Br{\"u}ggen} M., {Hoeft} M., {Oonk} J.~B.~R., 2012, \aap, 546, A124

\bibitem[{{Venturi} {et~al}\mbox{.}(2007){Venturi}, {Giacintucci}, {Brunetti},
  {Cassano}, {Bardelli}, {Dallacasa}, \& {Setti}}]{2007A&A...463..937V}
{Venturi} T., {Giacintucci} S., {Brunetti} G., {Cassano} R., {Bardelli} S.,
  {Dallacasa} D., {Setti} G., 2007, \aap, 463, 937

\bibitem[{{Venturi} {et~al}\mbox{.}(2008){Venturi}, {Giacintucci}, {Dallacasa},
  {Cassano}, {Brunetti}, {Bardelli}, \& {Setti}}]{2008A&A...484..327V}
{Venturi} T., {Giacintucci} S., {Dallacasa} D., {Cassano} R., {Brunetti} G.,
  {Bardelli} S., {Setti} G., 2008, \aap, 484, 327

\bibitem[{{Vikhlinin}, {Markevitch} \& {Murray}(2001){Vikhlinin}, {Markevitch},
  \& {Murray}}]{2001ApJ...549L..47V}
{Vikhlinin} A., {Markevitch} M., {Murray} S.~S., 2001, \apjl, 549, L47

\bibitem[{{Williams} {et~al}\mbox{.}(2012){Williams}, {Hewitt}, {Levine}, {de
  Oliveira-Costa}, {Bowman}, {Briggs}, {Gaensler}, {Hernquist}, {Mitchell},
  {Morales}, {Sethi}, {Subrahmanyan}, {Sadler}, {Arcus}, {Barnes}, {Bernardi},
  {Bunton}, {Cappallo}, {Crosse}, {Corey}, {Deshpande}, {deSouza}, {Emrich},
  {Goeke}, {Greenhill}, {Hazelton}, {Herne}, {Kaplan}, {Kasper}, {Kincaid},
  {Koenig}, {Kratzenberg}, {Lonsdale}, {Lynch}, {McWhirter}, {Morgan},
  {Oberoi}, {Ord}, {Pathikulangara}, {Prabu}, {Remillard}, {Rogers}, {Anish
  Roshi}, {Salah}, {Sault}, {Udaya Shankar}, {Srivani}, {Stevens}, {Tingay},
  {Wayth}, {Waterson}, {Webster}, {Whitney}, {Williams}, \&
  {Wyithe}}]{2012ApJ...755...47W}
{Williams} C.~L. {et~al.}, 2012, \apj, 755, 47

\bibitem[{{Wright} {et~al}\mbox{.}(1994){Wright}, {Griffith}, {Burke}, \&
  {Ekers}}]{1994ApJS...91..111W}
{Wright} A.~E., {Griffith} M.~R., {Burke} B.~F., {Ekers} R.~D., 1994, \apjs,
  91, 111

\bibitem[{{Wright}(2006)}]{2006PASP..118.1711W}
{Wright} E.~L., 2006, \pasp, 118, 1711

\bibitem[{{Wyllie}(1969)}]{1969MNRAS.142..229W}
{Wyllie} D.~V., 1969, \mnras, 142, 229

\bibitem[{{ZuHone}, {Markevitch} \& {Brunetti}(2011){ZuHone}, {Markevitch}, \&
  {Brunetti}}]{2011MmSAI..82..632Z}
{ZuHone} J., {Markevitch} M., {Brunetti} G., 2011, MmSAI, 82, 632

\bibitem[{{ZuHone} {et~al}\mbox{.}(2013){ZuHone}, {Markevitch}, {Brunetti}, \&
  {Giacintucci}}]{2013ApJ...762...78Z}
{ZuHone} J.~A., {Markevitch} M., {Brunetti} G., {Giacintucci} S., 2013, \apj,
  762, 78

\end{thebibliography}

\end{document}